# Constraining Active Galactic Nucleus Jets with Spectrum and Core Shift: The Case of M87

Kouichi Hirotani,[1] Hsien Shang （尚賢）,[1] Ruben Krasnopolsky,[1] Satoki Matsushita,[1] Britton Jeter,[1] and Keiichi Asada[1]

[1]*Institute of Astronomy and Astrophysics, Academia Sinica, Taipei 106216, Taiwan*

## ABSTRACT

We analytically model stationary and axisymmetric active galactic nucleus jets, assuming energy conservation along each magnetic flux tube. Using very-long-baseline interferometry (VLBI) observations and published general relativistic magnetohydrodynamic simulations, we constrain the evolution of the bulk Lorentz factor, the magnetization parameter, and the magnetic field strength along the jet. We then infer the electron density, emission coefficient, and absorption coefficient at each point, and integrate the radiative transfer equation to compute the spectral energy distribution (SED) and the core shift of the synchrotron emission from the relativistic jet. Applying the method to the M87 jet, we find that the hot plasmas are injected at the altitude of seven Schwarzschild radii from the black hole (BH), that the M87 jet is likely composed of a pair plasma, and that the jet flowline geometry is quasi-parabolic as reported at much greater distances. Fitting the nonthermal fraction of the leptonic jet as a function of position, we also find that most of the radio photons are emitted within 1000 Schwarzschild radii from the BH. Although hadronic jets do not reproduce all the VLBI observations consistently in our model, we also discuss that their heavy mass allows a stronger magnetic field within the observational constraints, leading to an inverted SED in sub-millimeter wavelengths by the thermal emission from the jet base. It is therefore implied that contemporaneous observations of the M87 jet with Atacama Large Millimeter/submillimeter Array (ALMA) and VLBI could discriminate the jet composition and its collimation within the central 100 Schwarzschild radii.

## 1. INTRODUCTION

Within the central regions of active galaxies, accreting supermassive black holes launch strong jets, observed as narrow beams of relativistic plasmas. The origin and early acceleration of these jets remain one of the most significant unresolved questions in contemporary astrophysics. Among numerous active galactic nuclei (AGNs), Messier 87 (M87, Virgo A, NGC4486, or 3C274) becomes an ideal laboratory for the study of extra-galactic jets from low-accreting AGNs, because of its proximity with the distance 16.7 Mpc (Blakeslee et al. 2009), and because of its large black hole (BH)

Corresponding author: Kouichi Hirotani, Hsien Shang
hirotani@asiaa.sinica.edu.tw,shang@asiaa.sinica.edu.tw

mass, $(3.5-6.6) \times 10^9 M_\odot$ (Walsh et al. 2013; Gebhardt et al. 2011; Event Horizon Telescope Collaboration et al. 2019a).

The jet of M87 was the first extragalactic jet identified, discovered by Curtis more than a century ago (Curtis 1918). In radio frequencies, high-resolution Very-Long-Baseline Interferometry (VLBI) imaging of inner jet regions provides the best constraints for the jet formation theories. So far, it has been established that the M87 jet has a wide opening angle of $\sim 60°$ on sub-pc scales (Junor et al. 1999) before expanding quasi-parabolically out to the distance of $\sim 10^5 R_S$ from the core (Asada & Nakamura 2012), where $R_S \equiv 2GMc^{-2} = 2M$ denotes the Schwarzschild radius. What is more, a limb-brightened structure is found on pc and sub-pc scales at 15 GHz (Kovalev et al. 2007), 43 GHz (Ly et al. 2007), and 86 GHz (Hada et al. 2016; Lu et al. 2023). At 230 GHz, The Event Horizon Telescope (EHT) revealed the center of the M87 galaxy and reported a ring-like structure with a diameter approximately 5 times Schwarzschild radii ($R_S$) (Event Horizon Telescope Collaboration et al. 2019b). The ring-like structure was confirmed at a lower frequency, 86 GHz, by Lu et al. (2023), with a greater diameter $\approx 8.4 R_S$. In their work, the limb-brightened structure (with an additional central ridge emission structure) was found to be extending down to the projected distance of 0.1 mas from the core, which corresponds to the de-projected distance of $\approx 45 R_S$ from the ring center.

To investigate how such brightness patterns are produced in jets, theoretical models have been proposed using force-free or MHD models. Integrating the radiative transfer equation in over-pressured super-fast-magnetosonic jets, Fuentes et al. (2018) investigated the images of jets with transverse structure and knots with a large variety of intensities and separations. Then, assuming a ring-like distribution of emitting electrons at the jet base, Takahashi et al. (2018) investigated the formation of limb-brightened jets in M87, Mrk 501, Cyg A, and 3C 84, and showed that symmetric intensity profiles require a fast spin of the BH. Subsequently, Ogihara et al. (2019) examined a steady axisymmetric force-free jet model, and found that the fluid's drift velocity produces the central ridge emission due to the relativistic beaming effect, and that the strong magnetic field and high plasma density near the edge results in a brightened limb. Kramer & MacDonald (2021) applied a polarized radiative transfer and ray-tracing code to synchrotron-emitting jets, and found that the jet becomes limb-brightened when the magnetic field is toroidally dominated, and that the jet becomes spine-brightened when the field is poloidally dominated. More recently, examining the transverse structure of MHD jet models, and computing radiative transfer of synchrotron emission and absorption, Frolova et al. (2023) showed that triple-peaked transverse profiles constrain the fraction of emitting leptons in a jet.

In terms of energetics, these relativistic flows can be fueled by either the extraction of the BH's rotational energy (Blandford & Znajek 1977) or the rotational energy of the accretion flow (Blandford & Payne 1982). In the recent two decades, general relativistic (GR) magnetohydrodynamic (MHD) simulations successfully demonstrated that a Poynting-dominated jet can be driven by the former process, the so-called "Blandford-Znajek (BZ) process" (Koide et al. 2002; McKinney 2005; Tchekhovskoy et al. 2010, 2011). The BZ process is driven when the horizon-crossing magnetic field is dragged in the direction of the BH's spin by space-time curvature, while a meridional current is present in the ergosphere, where the BH's rotational energy is stored. The resulting Lorentz force generates a counter-torque on the event horizon, extracting the BH's angular momentum and rotational energy, which are then carried away in the form of torsional Alfven waves. Subsequently, the application of the GR particle-in-cell (PIC) technique to magnetically dominated BH magnetospheres reveals that the BZ process continues to operate, even when Ohm's law breaks down and the MHD approximation becomes invalid in a collisionless plasma. (Parfrey et al. 2019; Chen & Yuan 2020; Kisaka et al. 2020; Crinquand et al. 2021; Bransgrove et al. 2021; Hirotani et al. 2023).

It is possible that the extracted electromagnetic energy is transformed into the kinetic and internal energies of electron-positron pairs at some distance from the BH, eventually being emitted as radiation through synchrotron and inverse-Compton processes in the jet downstream. We thus semi-analytically examine the jet emission properties, connecting the jet-launching and jet-downstream regions, using our post-processing, un-polarized radiative-transfer equation integration code R-JET (Hirotani et al. 2025, hereafter H25). In the present paper, we focus on the spectral energy distribution (SED) and the core shift of the synchrotron emission from the jet, and demonstrate that these two quantities strongly constrain the parameters of the M87 jet.

In § 2.1, we propose a method to infer the plasma density in a stationary jet, assuming energy conservation along each flow line, and evaluating the initial energy flux by the BZ process in the horizon vicinity. Then utilizing the published GRMHD simulations, and applying the method to the M87 jet, we describe the evolution of physical quantities along the jet, and quantify the emission and absorption properties at each position in § 3. Integrating the radiative transfer equation, we examine the SED and core shift for leptonic jets in § 4, and for hadronic jets in § 5. Finally in § 6, we consider implications to ALMA observations.

## 2. PARABOLIC JET MODEL

In the present paper, we consider stationary and axisymmetric jet. We assume that the jet is predominantly composed of electron-positron pair plasmas with a contamination by electron-proton normal plasmas. For electrons and positrons, which are referred to as leptons in the present paper, we consider both thermal and non-thermal components. In this section, we briefly summarize the method adopted in the R-JET code, whose details are described in our code paper (H25).

### 2.1. Poynting flux near the Black Hole

To model the angle-dependent BZ flux, we follow the method described in § 2 of H25 and Hirotani et al. (2024, hereafter H24). We model the flow lines of an axisymmetric jet with a parabolic-like geometry, adopting the magnetic-flux function (Broderick & Loeb 2009; Takahashi et al. 2018)

$$A_\varphi = A_{\max} \left(\frac{r}{r_{\rm H}}\right)^q (1 - |\cos\theta|), \quad (1)$$

where $r_{\rm H} = M + \sqrt{M^2 - a^2}$ denotes the horizon radius in geometrized units (i.e., $c = G = 1$), and $\theta$ denotes the colatitude; $c$ and $G$ refer to the speed of light and the gravitational constant, respectively. For the M87 jet, a quasi-parabolic flow-line geometry, $q = 0.75$, is reported within the de-projected altitude $10^{1.2} < r/R_{\rm S} < 10^{5.3}$ (Asada & Nakamura 2012; Asada et al. 2016; Nakamura et al. 2018). We thus assume a constant value, $q = 0.75$, in the present paper. We define the jet boundary by the magnetic flux surface that touches the horizon on the equator; in this case, we obtain $A_\varphi = A_{\max}$ along the jet boundary. Accordingly, the jet boundary position is given by $(r/r_{\rm H})^q(1 - \cos\theta) = 1$ on the poloidal plane, where we neglect GR corrections in the R-JET code.

Differentiating equation (1), and assuming a constant $q$ in the jet, we obtain the strength of the magnetic field,

$$B_{\rm p}(r,\theta) = B_{\rm p,0} \left(\frac{r}{r_0}\right)^{q-2} \sqrt{1 + q^2 \left(\frac{1-\cos|\theta|}{\sin\theta}\right)^2}. \quad (2)$$

In general, $r_0$ can be arbitrarily chosen. In the same way as H25, we set $r_0 = R_{\rm S}$ in this paper. Thus, $B_{\rm p,0}$ corresponds to the magnetic-field strength at this altitude. The value of $B_{\rm p,0}$ is adjusted so that the computed SED and core shift may match the observations (§ 4).

In a stationary and axisymmetric magnetosphere, the Poynting flux is given by

$$T^r{}_t = \frac{1}{4\pi}\left[F^{\mu\alpha}F_{\alpha t} + \frac{1}{4}g_t^\mu F^{\alpha\beta}F_{\alpha\beta}\right] = \frac{1}{4\pi}F^{r\theta}F_{\theta t}, \quad (3)$$

where $F_{\mu\nu}$ denotes the Faraday tensor. In the Boyer-Lindquist coordinates, we obtain the covariant component of the magnetic-field four vector:

$$B_\varphi = -\Sigma \sin\theta F^{r\theta}, \quad (4)$$

where $\Sigma \equiv r^2 + a^2\cos^2\theta$; $r$ and $\theta$ denotes the radial coordinate and the colatitude in the polar coordinates. The BH's spin parameter $a$ becomes $a = 0$ (or $a = M$) for a non-rotating (or an extremely rotating) BH; $M$ refers to the BH mass.

In ideal MHD, the frozen-in condition gives the meridional electric field, $F_{\theta t} = -\Omega_{\rm F} F_{\theta\phi}$, where $\Omega_{\rm F}$ denotes the angular frequency of rotating magnetic field lines. The radial component of the magnetic field is obtained by

$$\tilde{B}^r = \frac{A_{\max}}{\Sigma}\left(\frac{r}{r_{\rm H}}\right)^q, \quad (5)$$

where equation (1) is used. Thus, at the horizon, we obtain $B_{\rm p} \approx \tilde{B}^r = A_{\max}/\Sigma$.

We can constrain the meridional distribution of $F^{r\theta} = -B_\varphi/\Sigma\sin\theta$ in equation (4), using MHD simulations in the literature. For an enclosed electric current $I(A_\varphi)$ in the poloidal plane, we obtain

$$B_\varphi \approx -I(A_\varphi)/2\pi. \quad (6)$$

By GRMHD simulations, Tchekhovskoy et al. (2010) derived the current density

$$I(A_\varphi) \approx 6\sin\left[\frac{\pi}{2}(1-\cos\theta)\right](\omega_{\rm H} - \Omega_{\rm F})B_{\rm p} \quad (7)$$

in the poloidal plane. We thus obtain

$$B_\varphi \approx -\frac{3}{\pi}\sin\left[\frac{\pi}{2}(1-\cos\theta)\right](\omega_{\rm H} - \Omega_{\rm F})B_{\rm p}. \quad (8)$$

Combining equations (3), (4), (5), and (8), we obtain the angle-dependent BZ flux,

$$T^r{}_t \approx \frac{3}{4\pi^2}\sin\left[\frac{\pi}{2}(1-\cos\theta)\right]\Omega_{\rm F}(\omega_{\rm H} - \omega_{\rm H})B_{\rm p,H}^2, \quad (9)$$

where $B_{\rm p,H} \equiv B_{\rm p}(r = r_{\rm H})$. It follows that the BH's rotational energy is preferentially extracted along the magnetic field lines threading the horizon in the lower latitudes, $\theta \sim \pi/2$. The value of $B_{\rm p,H}$ can be readily computed by equation (2).

### 2.2. Kinetic flux inferred by the BZ flux

In this subsection, we connect the BZ flux inferred near the horizon (§ 2.1) with the kinetic flux in the jet downstream. To this end, we describe the kinetic flux in terms of bulk Lorentz factor and the co-moving energy density.

Considering that the jet plasma is composed of an electron-positron pair plasma and an electron-proton normal plasma, we can describe the proper number density of pair-origin electrons by

$$n_{*,e}^{\rm pair} = f_{\rm p} n_{*,e}^{\rm tot}, \quad (10)$$

where $n_{*,e}^{\rm tot}$ represents the density of electrons of both origins, and $f_{\rm p}$ does the fraction of pair contribution in number; $n_*$ denotes the number density in the jet co-moving frame. A pure *pair* plasma is obtained by $f_{\rm p} = 1$, whereas a pure *normal* plasma by $f_{\rm p} = 0$.

What is more, we assume that the leptons consist of thermal and nonthermal components, and express the nonthermal and thermal electron densities as

$$n_{*,e}^{nt} = w_{nt} n_{*,e}^{tot}, \quad (11)$$

and

$$n_{*,e}^{th} = (1 - w_{nt}) n_{*,e}^{tot} \quad (12)$$

respectively, where $w_{nt}$ denotes the nonthermal fraction of leptons.

In the BH-rest frame, the kinetic flux can be expressed as

$$F_{kin} = \beta c \Gamma (\Gamma - 1) n_{*,e}^{tot} U, \quad (13)$$

where $\beta c$ denotes the fluid velocity, $\Gamma \equiv 1/\sqrt{1-\beta^2}$ the bulk Lorentz factor. $U$ denotes the fluid mass per electron and is expressed as

$$U \equiv (1 + f_p) m_e c^2 \cdot \left[ \frac{3}{2} \Theta_e (1 - w_{nt}) + \langle \gamma \rangle w_{nt} \right] \\ + (1 - f_p) m_p c^2. \quad (14)$$

where $\Theta_e \equiv kT_e/m_e c^2$ denotes the dimensionless temperature of thermal leptons, $\langle \gamma \rangle$ the mean Lorentz factor of randomly moving nonthermal leptons, $m_e c^2$ the electron's rest-mass energy, and $m_p c^2$ the proton's rest-mass energy. The existence of a normal plasma (i.e., $f_p < 1$) becomes important in the present analysis when the second line dominates the first line through the heavier proton mass.

Neglecting the energy dissipation in the jet (Celotti & Fabian 1993), we obtain the kinetic-energy flux

$$F_{kin}(r, \theta) = \frac{1}{1+\sigma} \frac{B_p(r, \theta)}{B_{p,0}} F_{BZ,H} \quad (15)$$

in the jet, where $\sigma$ denotes the magnetization parameter, $B_p(r, \theta)$ does the poloidal magnetic field strength, and $F_{BZ,H}$ does the Poynting flux at $r = R_S$. If the magnetic field geometry is conical near the BH, we obtain $F_{BZ,H} = T^r{}_t$ (eq. [9]). Since the magnetic field lines are more or less radial at the horizon due to the plasma inertia and the causality at the horizon, we evaluate $F_{BZ,H}$ with equation (9) even when $q \neq 0$ at the horizon for simplicity.

Note that the total energy is conserved along each magnetic flux tube,

$$\frac{F_{EM} + F_{kin}}{B_p} = \frac{F_{BZ,H}}{B_p(r_H)}. \quad (16)$$

In this case, the Poynting flux evolves as

$$F_{EM}(r, A_\varphi) = \frac{\sigma}{1+\sigma} \frac{B_p(r, A_\varphi)}{B_p(r_H, A_\varphi)} F_{BZ,H}(r_H, A_\varphi), \quad (17)$$

where $\sigma \equiv F_{EM}/F_{kin}$. Accordingly, the toroidal component of the magnetic field evolves by

$$B_{\hat\varphi} = \frac{\sigma}{1+\sigma} B_{\hat\varphi}(r_H). \quad (18)$$

The more kinetically dominated the jet becomes, the more the toroidal magnetic field component declines.

### 2.3. Lepton energy distribution

Equations (13) and (15) show that electron density, $n_{*,e}^{tot}$, can be computed if $\Gamma$, $\sigma$, $f_p$, and $w_{nt}$ are specified at each position of the jet. We assume that the leptons are re-accelerated (e.g., in shocks) only within the altitude $r < r_2$, and adopt $r_2 = 3200 R_S$. In H25, we adopted $r_2 = 800 R_S$. However, in the present paper, to reproduce the observed flux densities below 10 GHz, we adopt $r_2 = 3200 R_S$ and pick up the synchrotron emission from the jet downstream at $r > 800 R_S$.

At $r < r_2$, we adopt the following form of the nonthermal lepton energy distribution,

$$\frac{dn_{*,e}^{nt}}{d\gamma} = (p-1) n_{*,e}^{nt} \gamma^{-p}, \quad (19)$$

where $\gamma$ denotes the Lorentz factor associated with their random motion, and $p$ the power-law index. The nonthermal lepton density $n_{*,e}^{nt}$ is specified by $w_{nt}$ (eq. [11]), once $n_{*,e}^{tot}$ is constrained by equations (13), (14), and (15). At $r > r_2$, the distribution function (eq. [19]) evolves by adiabatic expansion and synchrotron emission. For further details on these cooling processes, see § 2.3 of H25.

For thermal leptons, we assume that they have a relativistic temperature $kT = \Theta_{e,1} m_e c^2$ when they are injected in the jet. By adiabatic expansion, the lepton temperature $\Theta_e$ evolves as

$$\Theta_e = \Theta_{e,1} \left( \frac{B_p}{B_{p,1}} \right)^{\gamma_{ad}-1}, \quad (20)$$

where $\Theta_{e,1}$ denotes the temperature at the jet-injected altitude, $r = r_1 = 6.8 R_S$, where $B_{p,1}$ is evaluated. For the adiabatic index, we set $\gamma_{ad} = 4/3$. The relationship between $B_{p,0} \equiv B_p(R_S, 0)$ and $B_{p,1} \equiv B_p(6.8 R_S, 0)$ can be readily computed by equation (2).

We assume an energy equipartition between the energy density of (thermal+nonthermal) pairs and that of the random magnetic field, $B_{*,ran}^2/8\pi$. See § 2.3 of H25 for details. The strength of the *total* magnetic field is evaluated by $B = \sqrt{B_p^2 + B_{\hat\varphi}^2 + B_{*,ran}^2}$ at each position in the computation of the synchrotron process.

### 2.4. Emission and absorption coefficients

We assume the Maxwell-Jüttner distribution for thermal leptons, and a power-law distribution (eq. [19]) for nonthermal leptons. Once the distribution functions of thermal and nonthermal leptons, as well as the magnetic-field distribution are specified, we can compute the emission and absorption coefficients at each point in the jet. For a detailed method how to evaluate the emission and absorption coefficients, see § 2.4 of H25.

Using the emission and absorption coefficient at each point, we can integrate the radiative transfer equation along our line of sight to find the total specific intensity, where we neglect any polarization properties in the present paper. We can then compute the distribution of the surface brightness on the celestial plane to depict the expected VLBI map at each radio frequency, and compute the SED and the core shift as a function of frequency. These tasks can be done by the R-JET code.

## 3. STRUCTURE OF THE M87 JET

In the present paper, we apply the R-JET code to the M87 jet. We adopt the BH mass $6.4 \times 10^9 M_\odot$ (Gebhardt et al. 2011; Event Horizon Telescope Collaboration et al. 2019c; Liepold et al. 2023) and the luminosity distance 16.7 Mpc. An angular scale of 1 mas corresponds to a spatial scale $140 R_S$. We assume a rapidly rotating BH and adopt $a = 0.9M$ (Li et al. 2009; Dokuchaev 2023; Feng & Wu 2017).

### 3.1. Evolution of the bulk Lorentz factor

Investigating the proper motion of the components in the M87 jet, Asada et al. (2014) found that the jet exhibits a systematic increase of velocity from sub- to super-luminal speeds from mas to arcsecond scales. Subsequently, Mertens et al. (2016) obtained a detailed two-dimensional velocity field at sub-parsec scales, applying the wavelet-based image segmentation and evaluation (WISE) method to the M87 jet. Furthermore, performing ideal MHD simulations, they explained the obtained evolution of the flow velocity, and showed that the bulk Lorentz factor increases from $\Gamma \approx 2$ at distance $\varpi \approx 0.4$ mas from the jet axis (i.e., at de-projected distance $r \approx 140 R_S$ from the BH) to $\Gamma \approx 4.5$ at $\varpi \approx 1.5$ mas (i.e., at $r \approx 2800 R_S$). It should be noted that $r$ in Mertens et al. (2016) designates the distance $\varpi$ from the jet axis in the present notation. We thus approximate the evolution of the bulk Lorentz factor by

$$\Gamma = 2 + 2 \min \left[ \log_{10} \left( \frac{r}{150 R_S} \right), 2 \right] \quad (21)$$

as a function of the de-projected altitude $r$ in the present paper. At smaller scales, $r \ll 150 R_S$, we set a lower limit such that $\Gamma = \max(\Gamma, 1.2)$, whose actual value (e.g., 1.2 or 1.1) does not affect the results. At much larger scales, $r > 10^4 R_S$, the bulk Lorentz factor is found to be $\Gamma \sim 6$ by HST observations (Biretta et al. 1999). We thus set an upper limit of 6 in the right-hand side. Nevertheless, the saturation of $\Gamma$ at 6 does not play a role in the present analysis, because most emissions come from much inner regions, $r \ll 10^4 R_S$.

### 3.2. Evolution of the magnetization parameter

Assuming a quasi-parabolic jet flow line to the M87 jet, Mertens et al. (2016) applied ideal MHD simulations and found that the magnetization parameter $\sigma$, the ratio between the Poynting and kinetic energy fluxes, is much greater than 1.0 near the BH, but decreases with distance from the BH to become 2.0 at fast-magnetosonic point, whose projected distance is $z \equiv r \sin \theta_v \approx 1$ mas along the jet from the BH, where $\theta_v$ denotes the observer's viewing angle of the jet.

In the present paper, to explore a wide parameter space, we assume that $\sigma$ evolves with $r$ by a power law,

$$\sigma(r) = \sigma_0 \left( \frac{r}{100 R_S} \right)^{\sigma_p}. \quad (22)$$

Outside the fast point, $\sigma(r)$ continuously decreases and attain $\sigma \ll 1$, which means that the jet becomes kinetic-dominated asymptotically (e.g., Chiueh et al. 1998). We can mimic the $\sigma$ evolution of Mertens et al. (2016) by setting $\sigma_0 \approx 2.2$ and $\sigma_p \approx -0.25$. However, in our current jet model, this parameter set failed to reproduce the core shift as a function of frequency $\nu$, although we can obtain good SED fits when $\sigma_p = -0.25$. We thus loosen this constraint and adopt other combinations of $\sigma_0$ and $\sigma_p$. In the present paper, we adopt $\sigma_0 = 0.25$ and $\sigma_p = -1.0$ as the fiducial values, because this combination reproduces the observed SED and core shift well in our jet model.

It follows from equation (22) that the jet turns into kinetic-dominated around $r \approx 25 R_S$, provided that $\sigma_0 = 0.25$ and $\sigma_p = -1.0$. This rapid conversion of energy (from Poynting to kinetic) is, indeed, consistent with high-energy observations. For example, an efficient gamma-ray emission is reported during the VHE flares in year 2008 and 2010 (Acciari et al. 2009; Aliu et al. 2012; Abramowski et al. 2012). If this VHE emission is due to the synchrotron self-Compton process, it indicates the jet already becomes particle-dominated in the gamma-ray emitting region (Blandford & Levinson 1995). In addition, the VHE gamma-rays are appeared to be emitted within $20 R_S$ from the black hole (Hada et al. 2013). In this context, such a rapid conversion of energy within the central mas scales is observationally supported, although it has not been demonstrated by any theoretical simulation so far.

### 3.3. Fiducial case

Let us start with a parameter set that is consistent with observations. To this end, we assume that the magnetization parameter evolves with $\sigma_p = -1.0$, $\sigma_0 = 0.25$, that the ordered magnetic field has a strength $B_{p,0} = 25$ G and evolves with a quasi-parabolic shape, $q = 0.75$. In the lower latitudes, the magnetic field has comparable strengths in the poloidal and toroidal directions at the event horizon. Thus, the magnetic field strength becomes $B \approx 35$ G at $r = R_S$. It is worth noting that this value is within the observational constraints, which generally give $B < 100$ G for the M87 jet e.g., with the synchrotron self-absorption (SSA) theory (Kino et al. 2014, 2015; Hada et al. 2012, 2016; Reynolds et al. 1996; Ro et al. 2023).

For thermal leptons, we adopt $\Theta_{e,0} = 2.0$ as a semi-relativistic lepton temperature. For the nonthermal leptons, we set the power-law index $p = 3.1$, which gives the spectral index $\alpha = -1.05$ in the optically thin frequency regime.

We list their values in table 1. In § 4, we will show that this parameter set reproduces the observed SED and core shift of the M87 jet. In what follows, this set of parameters is referred to as the fiducial case.

As for the nonthermal fraction of leptons, we evaluate the value of $w_{\rm nt}$ at each $r$ by linearly interpolating its assumed values at discrete altitudes. In table 2, we tabulate such values of $w_{\rm nt}$ specified at 14 discrete altitudes. The second row shows the $w_{\rm nt} = w_{\rm nt}(r)$ that is commonly used in the present paper as the fiducial case. The third, fourth, and fifth rows will be used in § 4.3 to show how the SED and the core shift depends on the nonthermal fraction $w_{\rm nt}$ as a function of altitude $r$.

### 3.4. Distribution of magnetic field

In the following three subsections (§ 3.4–3.6), we describe the spatial distribution of physical quantities in the jet.

Let us begin with the evolution of the large-scale, ordered magnetic field as a function of altitude $r$ from the BH. In the poloidal plane, magnetic field strength is proportional to $r^{q-2}$ (eq. [2]). We adopt $q = 0.75$ as the fiducial value in this paper. Figure 1 shows the variation of $B_{\rm p}$ as a function of $r$ along five discrete magnetic flux tubes. The red solid curve corresponds to the magnetic-field line with $A_\varphi = 0.968 A_{\rm max}$, which is very close to the jet edge, $A_\varphi = A_{\rm max}$. The left panel shows a close-up view near the BH, whereas the right panel does a global view. Away from the BH, $B_{\rm p}$ little depends on $\theta$, because we have $|\theta| \ll 1$ in equation (2).

In the toroidal plane, on the other hand, magnetic field strength, $B_{\hat\varphi}$, decreases by $r^{-1}$ in ideal MHD well outside the Alfven point (Weber & Davis 1967). However, in general, mass loading (e.g., via pair production or entrainment) may be present in the jet. Such plasmas are accelerated outward by the Lorentz forces to attain the evolution of $\sigma$ as represented by equation (22). Because of this energy transfer from the electromagnetic field to the matter, the radial evolution of $B_{\hat\varphi}$ could deviate from the ideal MHD prediction, $B_{\hat\varphi} \propto r^{-1}$. Let us briefly consider this point below.

It follows from equation (18) that $B_{\hat\varphi} \propto r^{-1.0}$ at $\sigma \ll 1$ holds in the fiducial case, $\sigma \propto r^{\sigma_{\rm p}} = r^{-1.0}$. Although this $r$ dependence is coincidentally consistent with the ideal MHD prediction, a different power $\sigma_{\rm p}$ may be appropriate for a different jet in general.

In the left panel of figure 2, we show $\sigma$ as a function of $r$, adopting $\sigma_0 = 0.25$ and $\sigma_{\rm p} = -1.0$. Note that $\sigma$ has no dependence on $A_\varphi$ in the present model. Using this $\sigma = \sigma(r)$, we compute the evolution of $B_{\hat\varphi}$ as a function of $r$ along four discrete magnetic flux tubes (right panel). It is confirmed that $B_{\hat\varphi}$ does have $r^{-1.0}$ dependence in a kinetic-dominated jet, $\sigma \ll 1$, when $\sigma_{\rm p} = -1.0$.

Let us also see how $B_{\hat\varphi}$ depends on $A_\varphi$ (not $r$). In the left panel of figure 3, we show $|B_{\hat\varphi}|$ as a function of $A_\varphi$ at five discrete altitudes $r$, where the red solid, black dashed, green dash-dotted, blue dotted, and purple dash-dot-dot-dotted curves are evaluated at $r/R_{\rm S} = 10^{1.5}$, $10^{2.0}$, $10^{2.5}$, $10^{3.0}$, and $10^{3.5}$, respectively. The ordinate is in gauss, whereas the abscissa is normalized by $A_{\rm max}$. Thus, the abscissa 0 corresponds to the jet axis, whereas 1 does the jet limb. The values are plotted in the range $0 < A_\varphi/A_{\rm max} < 0.982$. It follows that the magnetic field is wound strongly away from the jet axis (i.e., towards the right side of this figure), as expected.

The right panel of figure 3 shows the strength of the random magnetic field, $B_{*,\rm ran}$, assuming an equipartition between the energy of random magnetic field and the internal energy of total leptons. We find that the random magnetic field (right panel) has less energy density than the ordered magnetic field near the jet limb, where the latter is dominated by the toroidal component (left panel). Thus, the toroidal component of the ordered magnetic field plays the primary role in the jet synchrotron emission in a kinetic-dominated jet (Vlahakis 2004; Lyubarsky 2009; Komissarov et al. 2009; Beskin & Nokhrina 2009; Beskin et al. 2023), whereas the random field plays a non-negligible but minor role.

### 3.5. Distribution of electron density

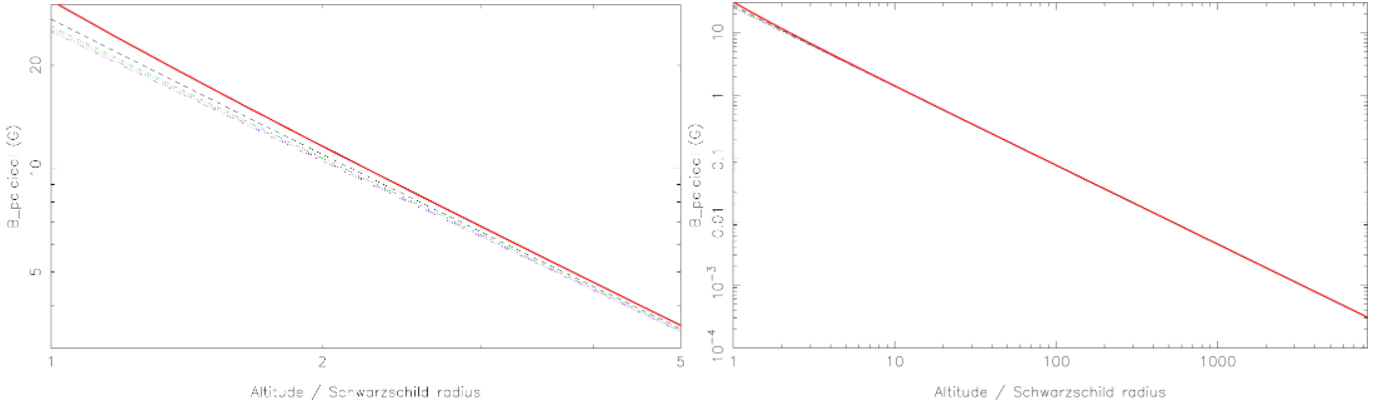

**Figure 1.** Radial variation of the magnetic field strength in the poloidal plane along five discrete magnetic flux tubes with $A_\varphi/A_{\rm max} = 0.03125$ (relatively near the jet axis, purple dash-dot-dot-dotted), 0.125 (blue dotted), 0.250 (green dash-dotted), 0.500 (black dashed), and 0.96875 (near the jet limb, red solid). *Left:* Zoom-in view near the event horizon. *Right:* Large-scale view away from the BH.

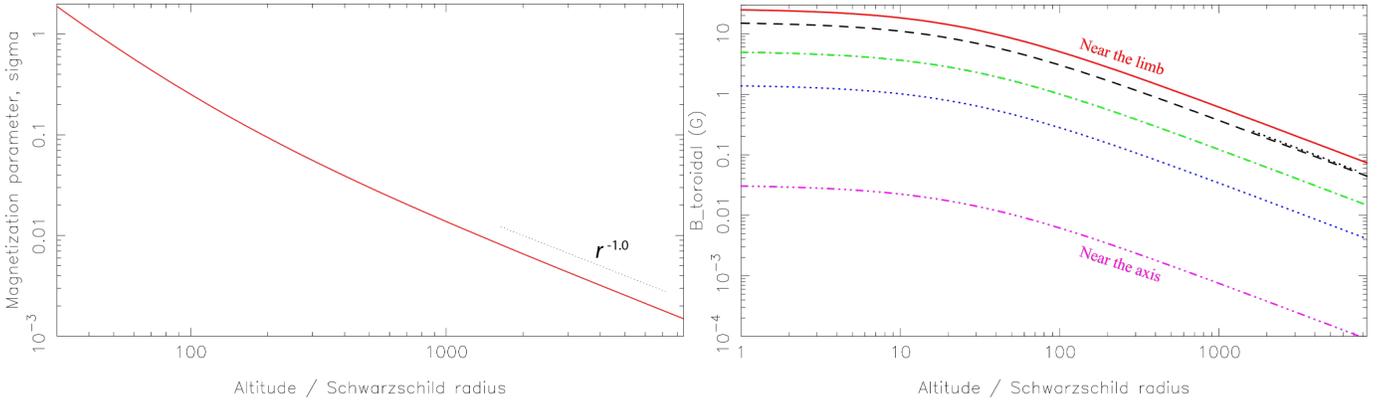

**Figure 2.** *Left:* Magnetization parameter as a function of dimensionless altitude, $r/R_{\rm S}$. For comparison, $r^{-1.0}$ dependence is overlaid as the thin black dotted line. *Right:* Strength of $B_{\hat\varphi}$ as a function of $r/R_{\rm S}$ along five discrete magnetic flux tubes with $A_\varphi/A_{\rm max} = 0.03125$ (purple dash-dot-dot-dotted), 0.125 (blue dotted), 0.250 (green dash-dotted), 0.500 (black dashed), and 0.96875 (red solid).

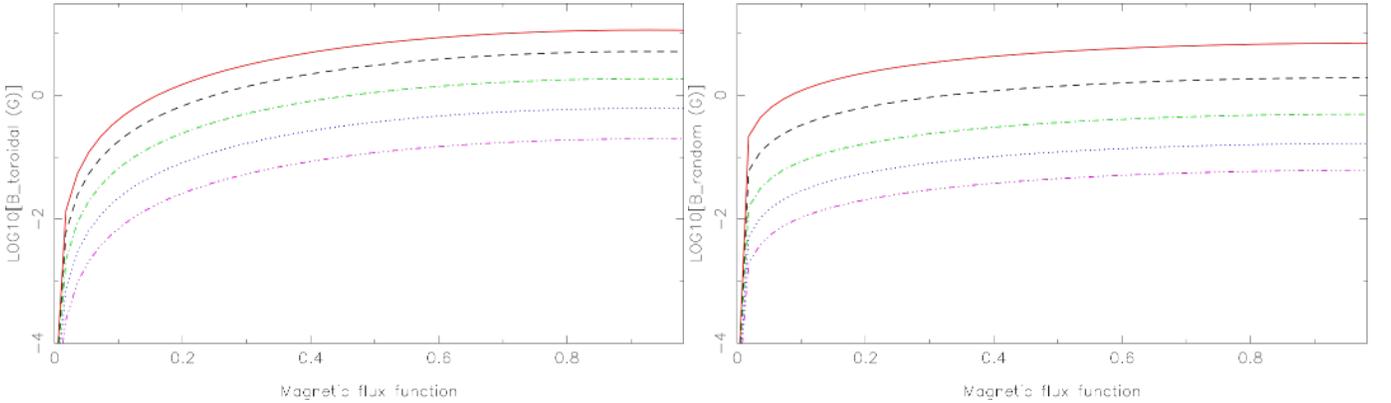

**Figure 3.** *Left:* Strength of the toroidal component of the magnetic field, $|B_{\hat\varphi}|$, as a function of the magnetic flux function $A_\varphi$. The red solid, black dashed, green dash-dotted, blue dotted, and purple dash-dot-dot-dotted curves are evaluated at the altitude $r/R_{\rm S} = 10^{1.5}$, $10^{2.0}$, $10^{2.5}$, $10^{3.0}$, and $10^{3.5}$, respectively. *Right:* Strength of the random magnetic field at the same positions in the poloidal plane $(r, A_\varphi)$ as the left panel.

Table 1. Fiducial model parameters for the M87 jet.

| Parameters | | Values | Comments |
|---|---|---|---|
| $q$ | Power-law index of magnetic-flux function, $A_\varphi$, on $r$ | 0.75 | Eqs. (1), (2) |
| $\sigma_{\rm p}$ | Power-law index of magnetization parameter, $\sigma$, on $r$ | $-1.0$ | Eq. (22) |
| $\sigma_0$ | Normalization of $\sigma$ | $0.25^*$ | Eq. (22) |
| $\Theta_{\rm e,1}$ | Dimensionless lepton temperature, $kT_{\rm e}/m_{\rm e}c^2$, at $r=r_1$ | 2.0 | Eq. (20) |
| $r_1$ | Altitude where jet plasmas are injected | $6.8R_{\rm S}{}^*$ | Eq. (2), (20) |
| $p$ | Power-law index of Lorentz factors of nonthermal leptons† | 3.1 | Eq. (19)–(??) |
| $\gamma_{\min}$ | Lower limit of Lorentz factors ($\gamma$) of nonthermal leptons† | $1.0^*$ | § 2.3 |
| $\gamma_{\max}$ | Upper limit of $\gamma$ at $r<r_2=3200R_{\rm S}$† | $10^{5\,*}$ | § 2.3 |
| $B_{\rm p,0}$ | Strength of poloidal magnetic field at $r=r_0=R_{\rm S}$ | 25 G | Eq. (2), (20) |
| $f_{\rm p}$ | Number fraction of pairs (1.0 means a pure pair plasma) | 1.0 | § 2.2 |

**Notes.** $^*$ means that the values are fixed in the fit. $^\dagger$ Random motion is concerned.

Table 2. Nonthermal fraction of pairs specified at discrete altitudes

| $r/R_{\rm S}$ | 6.8 | 12.5 | 25 | 50 | 100 | 200 | 400 | 800 | 1200 | 1600 | 2000 | 2400 | 2800 | 3200 | |
|---|---|---|---|---|---|---|---|---|---|---|---|---|---|---|---|
| $w_{\rm nt}$ | 0 | .05 | .05 | .05 | .05 | .05 | .03 | .01 | .01 | .05 | .10 | .50 | .90 | .90 | fiducial case |
| $w_{\rm nt}$ | 0 | 0 | 0 | 0 | 0 | .05 | .03 | .01 | .01 | .05 | .10 | .50 | .90 | .90 | red dashed (fig. 10) |
| $w_{\rm nt}$ | 0 | .05 | .05 | .05 | .05 | .05 | .03 | .01 | .01 | .05 | 0 | 0 | 0 | 0 | blue dotted (fig. 10) |
| $w_{\rm nt}$ | 0 | .05 | .05 | .05 | .05 | .05 | .03 | .01 | 0 | 0 | 0 | .50 | .90 | .90 | green dashed (fig. 11) |

Next, let us examine the electron density. In § 3, we assume a pure pair plasma and set $f_\mathrm{p}=1$; in this case, positrons have the same density as electrons. The left panel of figure 4 shows the number density of all (i.e., thermal plus nonthermal) electrons in the jet co-moving frame, $n_{*,\mathrm{e}}^\mathrm{tot}$, as a function of $A_\varphi$ at the five discrete altitudes. It is noteworthy here that the evolution of $n_{*,\mathrm{e}}^\mathrm{nt}$ is determined not only by the magnetic-field expansion, but also by the evolution of $\sigma$ along the jet. As a result of the angle-dependent energy extraction from the BH (§ 2.1), $n_{\mathrm{e},*}^\mathrm{tot}$ peaks in the jet limb, whose $A_\varphi$ is slight smaller than its boundary value, $A_\mathrm{max}$.

The right panel of figure 4 shows the density of nonthermal electrons, $n_{\mathrm{e}*}^\mathrm{nt}$. In the same way as the total number density (left panel), the nonthermal electrons have the greatest density at the jet limb at each altitude $r$.

Figure 5 show the evolution of $n_{\mathrm{e},*}^\mathrm{tot}$ (left panel) and $n_{\mathrm{e}*}^\mathrm{nt}$ (right panel) as a function of $r$ along five discrete magnetic flux function $A_\varphi$. Since the assumed nonthermal fraction $w_\mathrm{nt}$ has non-monotonic dependence on $r$ (table 2), $n_{\mathrm{e}*}^\mathrm{nt}$ distributes non-smoothly as the right panel shows. Accordingly, $n_{\mathrm{e},*}^\mathrm{tot}$ also shows a non-smooth $r$ distribution (left panel). In § 4, we will explain why such a functional form of $w_\mathrm{nt}(r)$ is needed for the observed SED and core shift to be reproduced.

### 3.6. Distribution of emission and absorption coefficients

Having the distributions of the magnetic field and the lepton densities are obtained at each point in the axisymmetric jet, we can now compute the emission and absorption coefficients. In this particular subsection (§ 3.6), for the purpose of demonstration, we adopt $\nu = 100$ GHz as the photon frequency in the observer's frame, and compute the absorption and emission coefficients in the jet co-moving frame, using Lorentz invariant relations (e.g., § 4.9 of Rybicki & Lightman 1986).

In the left and right panels of figure 6, we present the emission coefficients for thermal and nonthermal leptons, respectively, in cgs units (i.e., ergs s$^{-1}$cm$^{-3}$ster$^{-1}$Hz$^{-1}$). The red solid, black dashed, green dash-dotted, and purple dotted curves show the values measured along the magnetic flux tubes with $A_\varphi/A_\mathrm{max} = 1/4$, $1/2$, $3/4$, and $7/8$, respectively. The red solid curves in the left and right panels intersect at $r = 42.2 R_\mathrm{S}$, while the black dashed, green dash-dotted, and purple dash-dot-dot-dotted ones at $r = 89.8 R_\mathrm{S}$, $115.7 R_\mathrm{S}$, and $112 R_\mathrm{S}$, respectively. Therefore, the nonthermal emission dominates the thermal one near the jet limb (e.g., for the green and purple curves) at $r \gg 120 R_\mathrm{S}$. This is because thermal leptons are assumed to be produced at $r = 6.8 R_\mathrm{S}$, and cooled outside mostly by adiabatic expansion (and partly by synchrotron process), whereas the nonthermal leptons are assumed to be continuously energized (e.g., at shocks) at $r > 12.5 R_\mathrm{S}$ and be cooled by the adiabatic expansion and by the synchrotron process only in the outer region, $r > 3200 R_\mathrm{S}$.

Figure 7 shows the absorption coefficients in cm$^{-1}$ unit. The left and right panels are for the thermal and the nonthermal leptons. The four curves correspond to the absorption coefficients along the same magnetic-flux tubes in figure 6. The red solid, black dashed, green dash-dotted, and purple dash-dot-dot-dotted curves in the left and right panels, coincide at $r = 54.2 R_\mathrm{S}$, $115.7 R_\mathrm{S}$, $145.3 R_\mathrm{S}$, and $162.1 R_\mathrm{S}$, respectively. Therefore, the nonthermal absorption dominates the thermal one near the jet limb at $r \gg 160 R_\mathrm{S}$, whereas the absorption is mostly thermal inside this altitude.

Using the emission and absorption coefficients, we can integrate the radiative transfer equation to obtain the distribution of the surface brightness on the celestial plane. In H24, we present the surface brightness distribution and demonstrate that the jet exhibits a limb-brightened structure as a result of the angle-dependent BZ flux. Using the solved brighness distribution, we can compute the spectra of the entire jet, as well as the position of the brightness peak (i.e., the core) at each frequency on the celestial plane. In the next section, we focus on these two quantities: SED and core shift.

## 4. LEPTONIC JETS

### 4.1. Synchrotron spectrum: the fiducial case

To examine the SED, we begin with the fiducial set of parameters, which was also adopted in § 3 to show the distribution of the representative quantities (e.g., $B_{\hat{\varphi}}$, $n_{*,\mathrm{e}}^\mathrm{tot}$, $j_{*,\nu}^\mathrm{th}$, $j_{*,\nu}^\mathrm{nt}$, $\alpha_{*,\nu}^\mathrm{th}$, and $\alpha_{*,\nu}^\mathrm{nt}$) in the jet. In figure 8, we present the total SED as the black solid curve, where 'total' means that we take account of both the thermal and the nonthermal leptons as photon emitters. The black dashed one shows the flux density of the photons emitted by thermal leptons. The red solid, red dashed, red dash-dotted, red dotted, green solid, and green dashed curves denote the spectra of the photons emitted from the radial bins in $6.8 < r/R_\mathrm{S} < 10^1$, $10^1 < r/R_\mathrm{S} < 10^2$, $10^2 < r/R_\mathrm{S} < 10^3$, $10^3 < r/R_\mathrm{S} < 10^4$, $10^4 < r/R_\mathrm{S} < 10^5$, and $10^5 < r/R_\mathrm{S} < 10^6$, respectively. Filled symbols denote the flux densities observed during the low state. In radio frequencies, the filled circles show the flux densities of the M87 core with EVN, VERA, KVN, VLBA, or ALMA (Doeleman et al. 2012; Lonsdale et al. 1998; Lee et al. 2008; Junor & Biretta 1995; Morabito et al. 1986, 1988; Giovannini et al. 1990; Prieto et al. 2016;

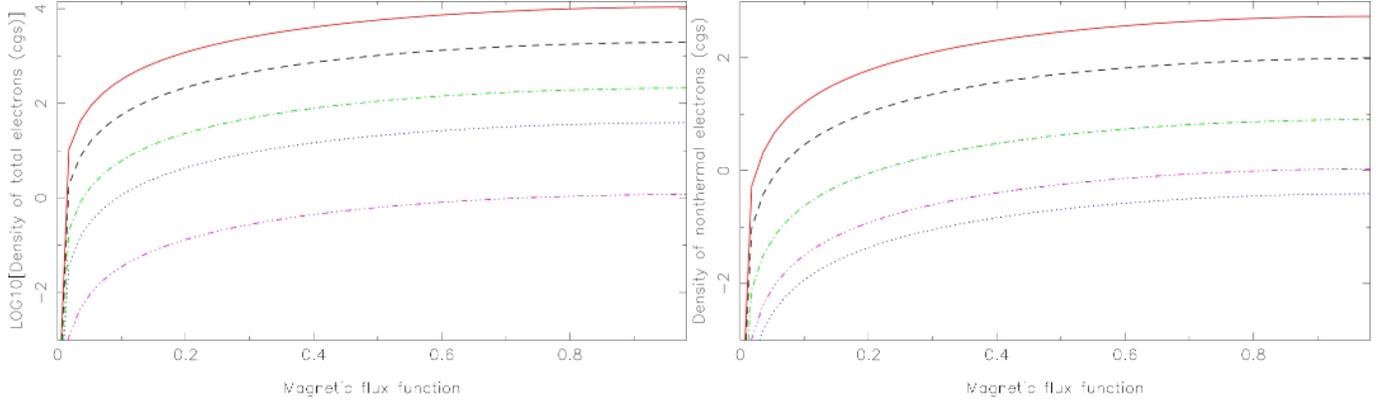

**Figure 4.** Proper number density of electrons as a function of $A_\varphi$, where the abscissa is common with figure 3. The five curves are measured at the same altitude $r$ as figure 3. *Left:* Density of electrons of all kinds. *Right:* Density of nonthermal electrons.

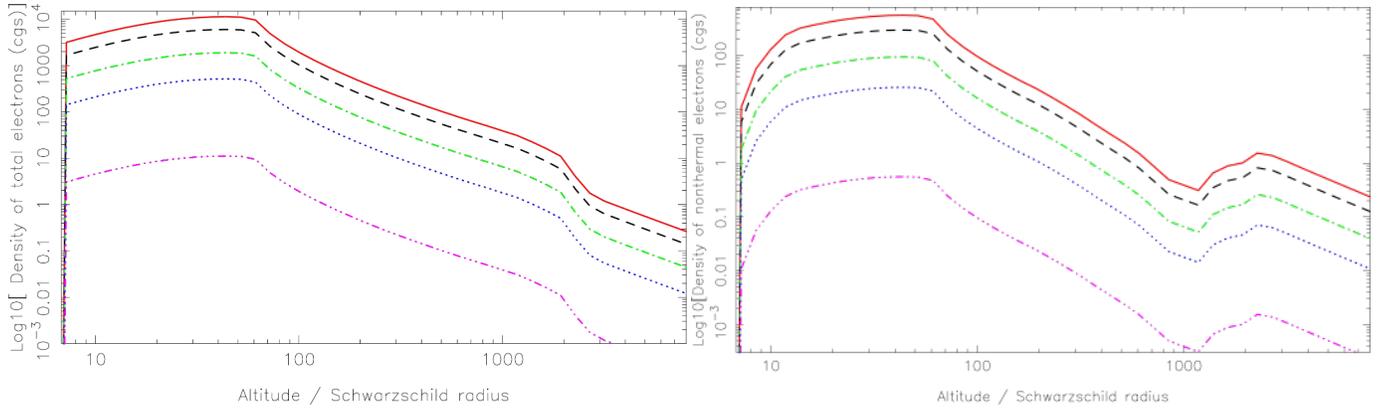

**Figure 5.** Proper number densities of electrons as a function of $r$, along five discrete magnetic flux tubes with $A_\varphi/A_{\max} =$ 0.03125 (purple dash-dot-dotted), 0.125 (blue dotted), 0.250 (green dash-dotted), 0.500 (black dashed), and 0.96875 (red solid). *Left:* For all the electrons. *Right:* For nonthermal electrons. Note the large difference in the ordinate scales between the two panels.

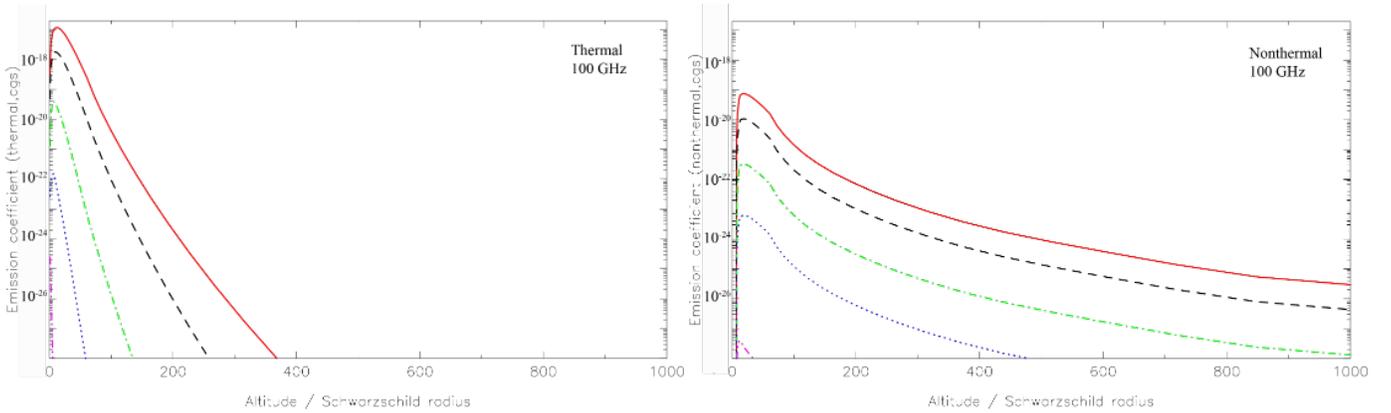

**Figure 6.** Emission coefficient in the jet co-moving frame as a function of $r/R_S$ along five discrete magnetic flux tubes with $A_\varphi/A_{\max} = 0.03125$ (purple dash-dot-dotted), 0.125 (blue dotted), 0.250 (green dash-dotted), 0.500 (black dashed), and 0.96875 (red solid). For demonstration purpose, we put $\nu = 100$ GHz as the photon frequency in the observer's frame. Values are plotted in the cgs unit, and the ordinates cover the same range in both panels. *Left:* Emission coefficient for thermal leptons (i.e., electrons and positrons), $j_{*,\nu}^{\mathrm{th}}$. *Right:* For nonthermal leptons, $j_{*,\nu}^{\mathrm{nt}}$.

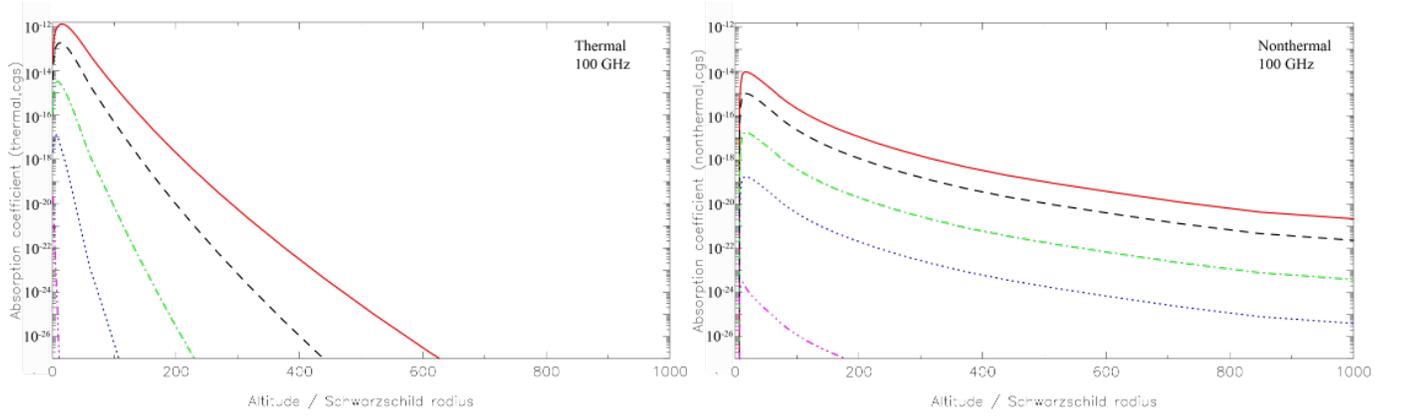

**Figure 7.** Absorption coefficient in the jet co-moving frame as a function of $r/R_S$ along five discrete magnetic flux tubes whose $A_\varphi$ takes the same values as figure 6. Values are computed for the frequency $\nu = 100$ GHz, and are plotted in unit of $cm^{-1}$. *Left:* Absorption coefficient for thermal leptons, $\alpha^{\rm th}_{*,\nu}$. *Right:* For nonthermal leptons, $\alpha^{\rm nt}_{*,\nu}$.

EHT MWL Science Working Group et al. 2021; Goddi et al. 2021; Peng et al. 2024; Goddi & Carlos 2025). The filled squares show the data in IR-optical-UV, obtained within an aperture radius of 0.15 arcsec in the HST data (Biretta et al. 1999). The filled triangles show the average of two Chandra epochs in year 2000 and 2002, with error bars showing the average variability of M87 core in X-rays (Harris et al. 2009). The optical and UV data give the upper limit to the core emission, $r \ll 0.15$ arcsec. In these higher frequencies (i.e., between IR and X-ray frequencies), data are selected according to the method described in Prieto et al. (2016). On the other hand, open symbols denote the flux densities during an active state, whose flux density increases by 10–35 % in 15–22 GHz comparing with the low state reported in Prieto et al. (2016). Open circles denote the flux densities obtained with VERA, EAVN, ALMA, SMA, and EHT, while open squares and triangles denote those obtained with HST, SWIFT, and Chandra, during the multi-wavelength campaign in year 2018 (Algaba et al. 2024, and references therein).

We briefly comment on the usage of the higher frequency ($\nu > 10^{14}$ Hz) observational data points. In the present paper, to focus on the SEDs, core shifts, and the radio maps in our kinematic jet modelling (§ 3), we do not incorporate either the free-free emission from the cold, standard Shakura-Sunaev accretion disk outside the truncation radius (within which an advection-dominated accretion flow appears), or the synchrotron self-Compton (SSC) emission from the relativistic jet. In this context, we regard the observed flux densities (open circles and filled triangles in fig. 8) as the upper limits. What is more, for simplicity, we ignore the X-ray absorption (by hydrogen and heavier elements) and the optical extinction (by grains composed of the same heavier elements). These two effects could loosens the range of parameters in our current treatment (in which the higher frequency observational data points are treated as the upper limits), without giving additional constraints. On the other hand, if the standard-disk and the jet-SSC emissions were incorporated, it would be worth incorporating the absorption and the opacity effects when we fit the spectrum (e.g., Prieto et al. 2016).

The left panel of figure 8 shows that the total spectrum (black solid curve) is in fair agreement with the radio observations in cm, mm, and sub-mm wavelengths. The figure also shows that the photons are mostly emitted from the altitudes within $10^1 R_S < r < 10^2 R_S$ at higher frequencies $\nu > 15$ GHz, as the red dashed curve shows. However, the SED is be kept relatively flat in lower frequencies $\nu < 15$ GHz, by the emission from the higher altitudes in $10^2 R_S < r < 10^4 R_S$, as the red dash-dotted and red dotted curves show. For the present semi-relativistic temperature (namely, $\Theta_e = 2.0$ at $r = 6.8 R_S$), the thermal component contributes in the SED only between 30 and 100 GHz, as the black dashed curve shows. At $r > 10^3 R_S$, photon spectrum decreases rapidly with distance $r$ (as the red dash-dotted, red dotted, green solid and green dashed curves show), because the magnetic field strength decreases with $r$.

The right panel of figure 8 shows that the core shift is also in good agreement with observations (Hada et al. 2011). As will be demonstrated in §§ 4.3–5.2, simultaneous fitting of SED and core shift enables significantly tighter constrains on various parameters (e.g., acceleration site of leptons, evolution of the $\sigma$ parameter, magnetic-field strength, the collimation index, as well as the matter composition) than fitting the SED alone.

### 4.2. Ring-like structure in leptonic jet

We next consider the expected VLBI map. Figure 9 shows the surface brightness distribution on the celestial plane for the fiducial case at photon frequency 230 GHz in the observer's frame. Because of the angle-dependent energy extraction from the BH (fig. 1 of H24), synchrotron emission from the jet base exhibits a ring-like structure. In our present model, we assume that thermal pairs are injected at $r = 6.8 R_S$, and nonthermal pairs at $r > 12.5 R_S$ with $w_{nt} = 0.05$. In this case, the thermal photons are emitted at $r > 6.8 R_S$, and form a ring-like structure with a diameter $64 \mu$as. The injection altitude is, indeed, assumed to be at $r = 6.8 R_S$ so that the ring diameter may be consistent with what was reported in Lu et al. (2023). Therefore, if this ring-like structure is formed by the photons emitted from the jet limb, as discussed by Lu et al. (2023), it indicates that the base of the M87 jet is located in a BH vicinity at $r \approx 6.8 R_S$.

### 4.3. Dependence on the nonthermal fraction

Let us return to the main topic of this paper and focus on the SED and the core shift. In this subsection, we investigatie how these quantities depend on the nonthermal fraction, $w_{nt} = w_{nt}(r)$. Although the distribution of $w_{nt}$ should be, in principle, determined by the particle re-acceleration processes taking place in shocks or reconnection sites, it is out of the scope of this paper to deduce the position of the lepton acceleration or the energy distribution of the accelerated leptons from the first principles. In the previous subsection § 4.1, we thus assumed the functional form of $w_{nt}(r)$, and examined the SED for the fiducial case.

In this subsection (§ 4.3), we next consider the impact when $w_{nt}$ changes from the fiducial case (second row of

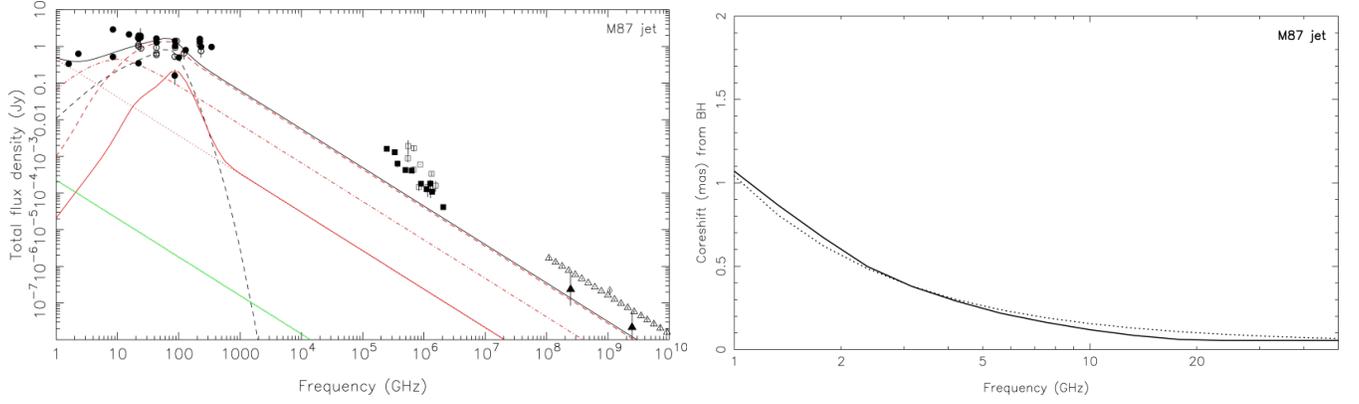

**Figure 8.** Spectral energy distribution (SED) and core shift of the M87 jet for the fiducial set of parameters: $\sigma_p = -1.0$, $\sigma_0 = 0.25$, $B_{p,0} = 25$ G, $q = 0.75$, $p = 3.1$, $\Theta_e = 2.0$. *Left:* The black solid curve shows the flux density of the photons emitted by all the thermal and nonthermal leptons, whereas the black dashed one does the thermal component. The red solid, red dashed, red dash-dotted, red dotted, green solid, and green dashed curves denote the spectra of the photons emitted from the radial bins in $6.8 < 5/R_S < 10^1$, $10^1 < r/R_S < 10^2$, $10^2 < r/R_S < 10^3$, $10^3 < r/R_S < 10^4$, $10^4 < r/R_S < 10^5$, $10^5 < r/R_S < 10^6$, respectively. For the observational data points, which are denoted by filled and open symbols, we follow the selection method described in Prieto et al. (2016). *Right:* The solid curve shows the core shift obtained by the R-JET code for the fiducial case, while the dotted one does the observational values (Hada et al. 2011).

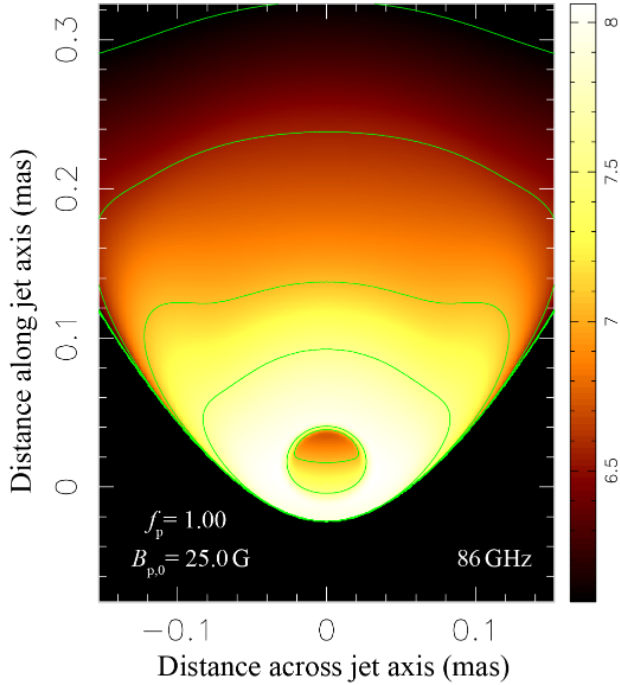

**Figure 9.** Surface brightness distribution of the jet core for the fiducial case. The origin (0,0) shows the line of sight toward the BH. The color code is in the logarithm of $\mu$Jy mas$^{-2}$ unit. The peak brightness is 115.3 Jy mas$^{-2}$. Photon frequency is 86 GHz in the observer's frame.

table 2). To this end, we consider three cases whose $w_{\rm nt}$ vanishes in the inner, outer, and middle parts of the jet. In the first two cases, we set $w_{\rm nt} = 0$ at $r \leq 100R_S$ and $r \geq 2000R_S$. In the third case, we set $w_{\rm nt} = 0$ at $1200R_S < r < 2000R_S$. These three cases are listed in the third, fourth, and fifth rows in table 2.

Figure 10 shows the results for the first two cases: the red dashed curve corresponds to the case of $w_{\rm nt}(r \leq 100R_S) = 0$, whereas the blue dotted one does $w_{\rm nt}(r \geq 2000R_S) = 0$, where the altitude $r = 2000R_S$ corresponds to the projected distance 4.76 mas on the celestial plane when $\theta_v = 18°$. Comparing the black solid and red dashed curves in the left panel, we find that the flux density increases at $\nu > 7$ GHz by an increase of nonthermal pairs in the inner part of the jet, $r < 100 R_S$. What is more, comparing the black solid and blue dotted curves, we find that the flux density increases at $\nu < 6$ GHz by an increase of nonthermal pairs at $r > 2000 R_S$. In this figure, as well as in figures 11–19, we plot only the observational flux densities reported in Prieto et al. (2016) to prevent confusion.

The right panel shows that the core shift is not affected by the emission from $r > 2000 R_S$, because the black solid and blue dotted curves coincide. It is noteworthy that $r = 2000 R_S$ corresponds to a projected distance of $z = 4.4$ mas along the jet axis for a highly collimated jet. (For a non-collimated jet, 3D projection effect results in a wider range of $z$.) Since both the magnetic field strength as well as the lepton number density is small at such a large distance, opacity is too small to affect the core shift. On the other hand, the nonthermal leptons in $r < 100 R_S$ do not strongly affect the core shift either, because the thermal emission and absorption dominate the nonthermal ones in this spatial region below 100 GHz. (§ 3.6). Accordingly, they little affect the core shift, as the black solid and red dahsed curves show in the right panel.

To examine the third case, we set $w_{\rm nt} = 0$ in the middle part, $1200 R_s < r < 2000 R_s$ (fifth row in table 2, where the intermediate values are linearly interpolated from the tabulated values). The green dashed curve in figure 11 shows the result of this case, whereas the red solid curve does the fiducial case. It follows that the lack of nonthermal emission within the altitude $1200 R_s < r < 2000 R_s$ little affects the SED or the core shift, as long as $w_{\rm nt}$ is smaller than 0.01, 0.05, and 0.10 at $r = 1200 R_S$, $1600 R_S$, and $2000 R_S$, respectively (table 2). However, if we increase $w_{\rm nt}$ than these values in $1200 R_S < r < 2000 R_S$, the core shift increases too much below 2 GHz, deviating from the observed values (dotted curve in the right panel of fig. 8).

In short, the SED and the core shift give strong constraints on theoretical models, or on model parameters. As long as the re-acceleration sites of nonthermal leptons are concerned, the re-acceleration in $r < 1200 R_s$ contributes to most of the emissions in higher frequencies ($\nu > 7$ GHz), whereas that in $r > 2000 R_s$ contributes to the formation a flat SED in lower frequencies ($\nu < 6$ GHz).

### 4.4. Dependence on the $\sigma$ evolution

Let us next examine the dependence on the evolution of the magnetization parameter, $\sigma$, with altitude. In figure 12, we present the results for the three cases of discrete power-law indices: $\sigma_{\rm p} = -0.5$, $-1.0$, and $-1.5$, where the black solid curve corresponds to the fiducial case. The left panel shows that the flux density increases with decreasing $|\sigma_{\rm p}|$. It is worth noting that a smaller $|\sigma_{\rm p}|$ indicates a slow mass loading, as well as a slow reduction of the toroidal compoent $|B_{\hat{\varphi}}|$ of the magnetic field. A slow mass loading indicates a slow increase of pair density, resulting in a smaller flux density of photons. On the other hand, a slow reduction of $|B_{\hat{\varphi}}|$ leads to a greater flux density due to synchrotron emission. These two effects cancel each other, resulting in a weak dependence of the SED flux density as a function of $\sigma_{\rm p}$. Thus, this cancellation effect results in a slight difference in both the thermal component around 60 GHz and the nonthermal, optically thin component above 200 GHz. A close examination reveals that the latter effect slightly dominates the former one, leading to an increasing flux density in lower frequencies with a decreasing power, $|\sigma_{\rm p}|$, by virtue of the emission from larger distances with greater $B_{\hat{\varphi}}$.

The right panel shows that the core shift rapidly increases when $|\sigma_{\rm p}| < 1.0$. This is because a greater $|B_{\hat{\varphi}}|$ (for a slower energy conversion from EM to kinetic) leads to a stronger synchrotron absorption by thermal pairs.

### 4.5. Dependence on the poloidal magnetic field

Let us examine the dependence on $B_{\rm p,0}$, the strength of the ordered magnetic field in the poloidal plane at $r = 6.8 R_S$. In figure 13, we compare the three cases: $B_{\rm p,0} = 20$ G, 25 G, and 30 G, where the black solid curve corresponds to the fiducial case. It follows that the flux density increases and the SED hardens with increasing $B_{\rm p,0}$, because both thermal and nonthermal emissions increase with increasing magnetic-field strength. It also follows that the core shift increases at each frequency with increasing $B_{\rm p,0}$, because synchrotron absorption (by thermal leptons in this case) takes place more efficiently for a stronger magnetic field.

Next, we examine the dependence on the *configuration* of the magnetic field lines in the poloidal plane. In figure 14, we plot the results for three discrete collimation indices: $q = 0.65$, 0.75, and 0.85. The left panel shows that the SED little depends on the poloidal field configuration. The reasons are twofold: The optically thin synchrotron emission takes place mainly slightly outside the injection point; namely at $r \approx 6.8 R_S$ for thermal leptons and $r \approx 12.5 R_S$ for nonthermal ones. Accordingly, the difference of $q$ gives a slight difference in both the lepton density and the magnetic-field strength near the injection point.

On the other hand, the core shift strongly depends on $q$, as the right panel shows. This is because the thermal and nonthermal absorption takes place in the further downstream of the injection points, thereby being affected by the radial evolution of the lepton density and the magnetic-field strength. As a result, the core shift increases with increasing $q$, because a more collimated jet results in a stronger magnetic field and a denser plasma density in the downstream.

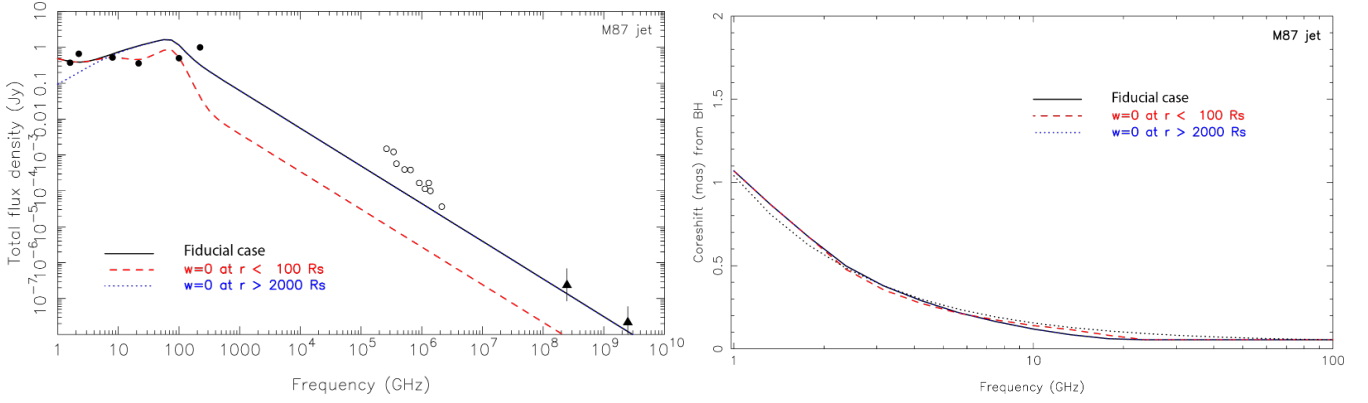

**Figure 10.** *Left:* Total SEDs of the M87 jet for three different cases of nonthermal fraction of pairs. The black solid curve corresponds to the case of figure 8, the fiducial case; thus, it is the same curve as the black solid one in fig. 8. The red dashed curve corresponds to the case when there is no nonthermal pairs at $r < 100R_{\rm S}$, whereas the blue dotted when there is no nonthermal pairs at $r > 2000R_{\rm S}$. The black solid and red dashed curves coincide below 4 GHz, whereas the black solid and blue dotted ones do above 4 GHz. Other parameters than $w_{\rm nt}(r)$ are common with the fiducial case (fig. 8). To prevent confusion, only the observational flux densities reported in Prieto et al. (2016) are plotted. *Right:* Core shifts for the same three cases. The thin black dotted curve represents the observed values (Hada et al. 2011).

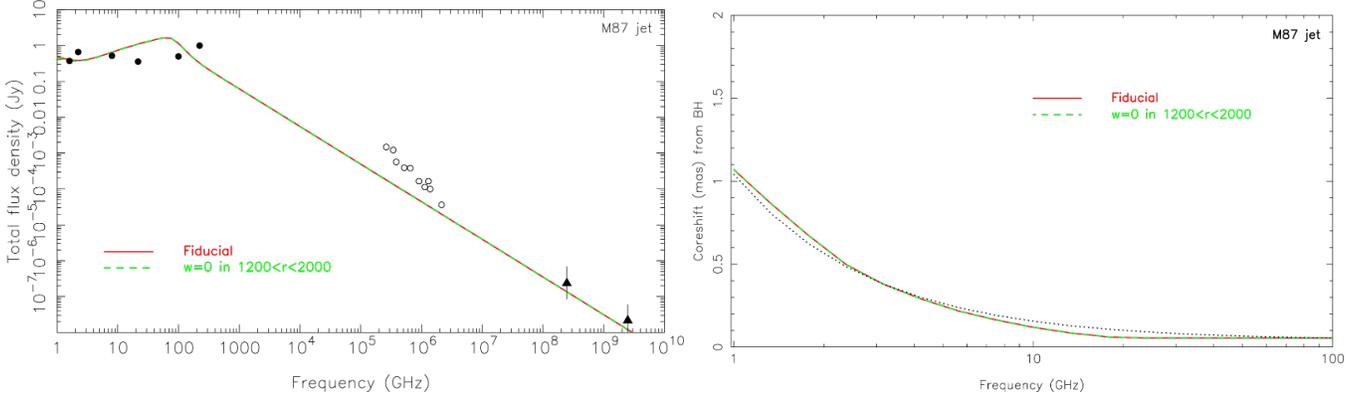

**Figure 11.** *Left:* Comparison of SED for the fiducial case and the case when $w_{\rm nt} = 0$ in $1200R_{\rm s} < r < 2000R_{\rm s}$ (black dashed). To show the overlapped curves, a red solid curve is used for the former case, while a green dashed one for the latter case. *Right:* Core shifts for the same two cases. The thin black dotted curve represents the observed values.

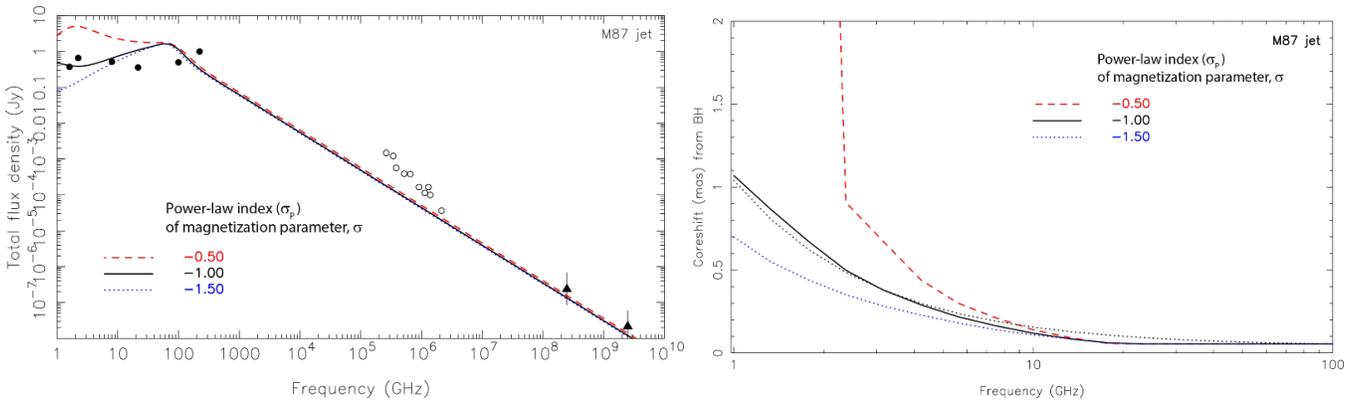

**Figure 12.** *Left:* Total SEDs for three different cases of $\sigma_{\rm p}$ (i.e., power of the evolution of the magnetization parameter, $\sigma$). The black solid curve corresponds to the fiducial case (fig. 8), that is, $\sigma_{\rm p} = -1.0$. The red dashed curve corresponds to $\sigma_{\rm p} = -0.5$, whereas the blue dotted one to $\sigma_{\rm p} = -2.0$. Other parameters are the same as the fiducial case. *Right:* Core shifts for the same three cases.

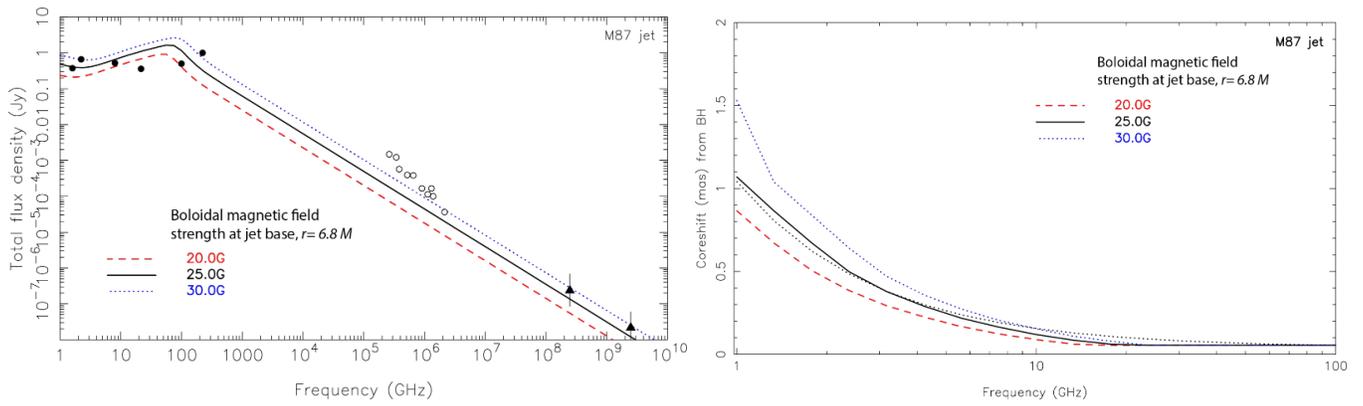

**Figure 13.** *Left:* Total SEDs for three different cases of $B_{p,0}$ (i.e., strength of magnetic field at $r = 6.8M$ in the poloidal plane). The black solid curve corresponds to the fiducial case (fig. 8), that is, $B_{p,0} = 25$ G, whereas the red dashed and blue dotted ones to 20 G and 30 G, respectively. *Right:* Core shifts for the same three cases.

### 4.6. *Dependence on the lepton temperature*

Figure 15 shows the SED and the core shift for three different temperatures, $\Theta_{e,0} = 1.0$, $2.0$, and $3.0$, where the black solid curve corresponds to the fiducial case. Since an increase of $\Theta_{e,0}$ results in a decrease of $n_{*,e}^{\rm tot}$ by equation (13), nonthermal emission decreases with increasing $\Theta_{e,0}$, as the left panel shows. The thermal component appears between 50 and 100 GHz, increasing the peak frequency with increasing $\Theta_e$. The reasons are threefold: (1) The peak frequency is determined by the condition $\tau_\nu \equiv \alpha_{*,\nu} l_* \approx 1$, where $\tau$ designates the optical depth and $l_*$ the typical length of the radio-emitting blob in the co-moving frame along the line of sight. Thus, to satisfy $\tau \approx 1$ for a fixed $l_*$, $\alpha_{*,\nu}$ should not change when $\Theta_e$ changes. (2) $\alpha_{*,\nu}$ strongly depends on $X \propto \nu_*/\Theta_e^2$ through the factor $\exp(-X^{1/3})$ (§ 2.4), where $X \ll 1$ in the present case. (3) To keep $X$ nearly constant, and hence $\tau \approx 1$ little changed, the peak frequency $\nu$ should increase when $\Theta_e$ increases.

The right panel shows that the core shift decreases with increasing $\Theta_e$, because the increased $U$ in equation (14) results in a smaller $n_{*,e}^{\rm tot}$. The smaller density leads to a decreased $\alpha_{*,\nu}$, and hence the core shift.

### 4.7. *Dependence on nonthermal power law*

In figure 16, we present results for three discrete power-law indices of nonthermal leptons: the red dashed, black solid, and blue dash-dotted curves show the SEDs and core shifts for $p = 2.9$, 3.1, and 3.3, respectively. As expected, a softer power law (in the present case, $p = 3.3$) of lepton energy distribution results in a softer synchrotron spectrum in the optically thin regime, $\nu > 10^3$ GHz. The spectral index in the optically thin regime becomes $(1-p)/2$, which is consistent with the standard argument for the synchrotron radiation from the electrons with a power-law energy distribution.

We next consider the core shift. The right panel shows that the gradient of the frequency dependence has only a weak dependence on $p$, as long as $p \approx 3$. Let us analytically confirm this conclusion, by considering only the power-law, nonthermal leptons for simplicity. Since the nonthermal leptons dominates in absorption at $r > 10^{2.5} R_{\rm S}$, the argument in the rest of § 4.7 is valid only in $r > 10^{2.5} R_{\rm S}$. Following Konigl (1981), we assume a power-law dependence of the lepton density and the magnetic field strength as

$$n_{*,e}^{\rm nt} \propto r^{-n} \qquad (23)$$

and

$$B = \sqrt{B_{\rm p}^2 + B_{\hat\varphi}^2} \propto r^{-m}. \qquad (24)$$

In this case, the core shift will depend on frequency $\nu$ by (Lobanov 1998)

$$r_{\rm core}(\nu) \propto \left(B^{k_b} n_*^{k_c}\right)^{1/k_r} \nu^{-1/k_r}, \qquad (25)$$

where

$$k_r = \frac{(p+2)m + 2n - 2}{p+4}, \qquad (26)$$

$$k_b = \frac{p+2}{p+4}, \qquad (27)$$

$$k_c = \frac{2}{p+4}. \qquad (28)$$

We thus obtain

$$k_r \approx m + \frac{2(n-m-1)}{7}\left(1 - \frac{p-3}{7}\right) \qquad (29)$$

In our present model, in $10^{2.5} R_{\rm S} < r < 10^3 R_{\rm S}$, $n \approx 1.5$ (right panel of fig. 5). For the magnetic field, we obtain $B \approx B_{\hat\varphi} \propto r^{-1.0}$, which gives $m \approx 1.0$. Accordingly, equation (29) gives

$$k_r \approx 0.857 + 0.0204(p-3) \qquad (30)$$

Therefore, as long as $|p-3| \ll 1$, we obtain $k_r \approx 0.857$. It is noteworthy that the value of $k_r$ is within the range $(0.6 < k_r < 1.1)$ reported by observations (Hada et al. 2011; Sokolovsky et al. 2011; Kutkin et al. 2014; Ricci et al. 2022; Nokhrina & Pushkarev 2024).

## 5. HADRONIC JETS

The particles injected into a jet may have their origin not only in a relativistic *pair* plasma, but also in a semi-relativistic *normal* plasma. The former, a pair plasma could be supplied by a pair-production cascade in a BH magnetosphere (Beskin et al. 1992; Hirotani & Okamoto 1998; Neronov & Aharonian 2007; Levinson & Rieger 2011; Broderick & Tchekhovskoy 2015; Hirotani & Pu 2016; Kisaka et al. 2020). On the other hand, the latter, normal plasma could be supplied by an advection from a RIAF, and consist of protons and electrons as long as helium and heavier elements are ignored. In this section, we thus investigate the impact of a hadronic contamination in a jet, assuming a pure-hydrogen plasma as the normal plasma. Note that we put $f_{\rm p} < 1$ in equation (14) when we consider the existence of protons in the jet.

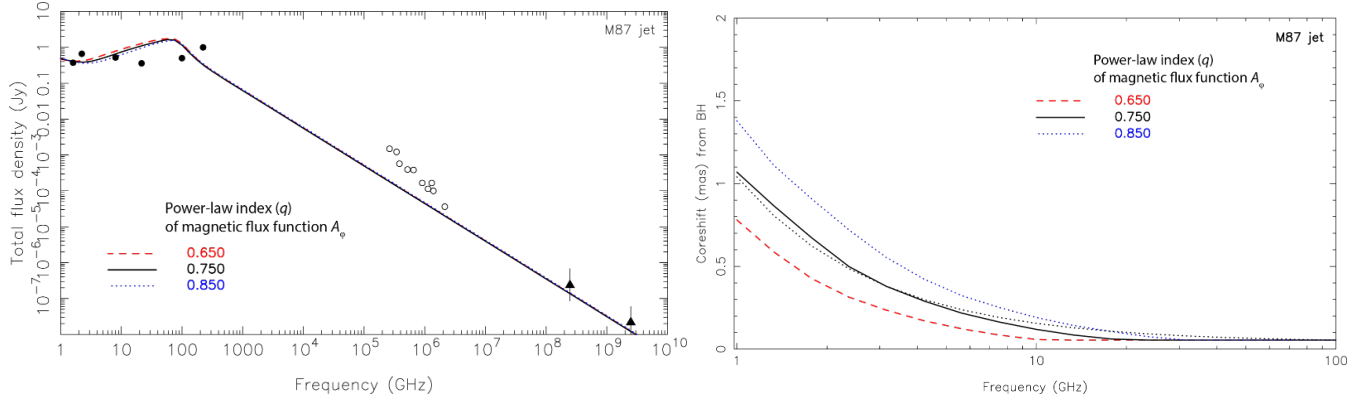

**Figure 14.** *Left:* Total SEDs for three different cases of the power-law index of $A_\varphi$, $q$. The black solid curve corresponds to the fiducial case (fig. 8), that is, $q = 0.75$, whereas the red dashed and blue dotted ones to $q = 0.65$ and $q = 0.85$, respectively. *Right:* Core shifts for the same three cases.

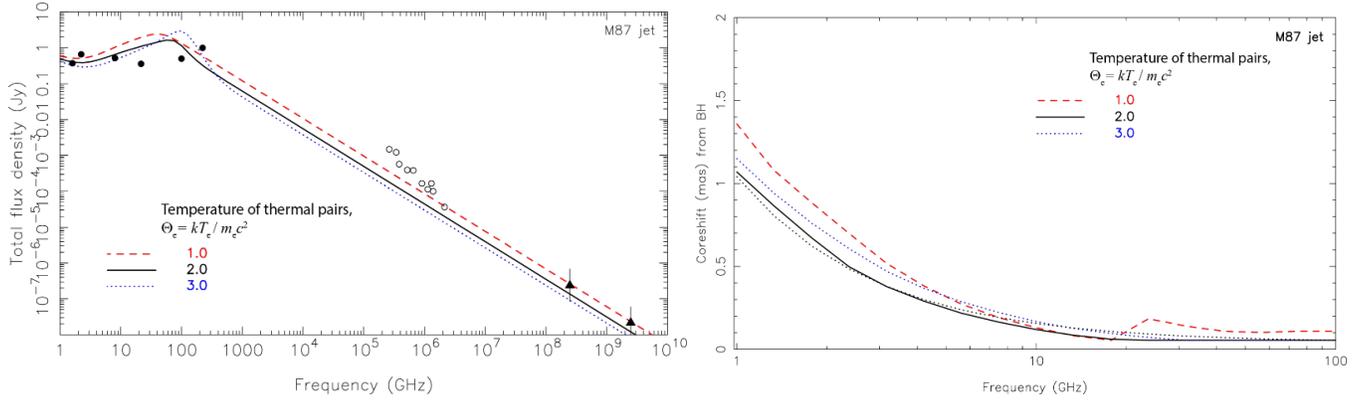

**Figure 15.** *Left:* Total SEDs for four different temperatures of thermal pairs, $\Theta_{e,0}$. The black solid curve corresponds to the fiducial case (fig. 8), that is, $\Theta_{e,0} = 2.0$, whereas the red dashed, blue dotted ones correspond to $\Theta_{e,0} = 1.0$ and $3.0$, respectively. *Right:* Core shifts for the same three cases.

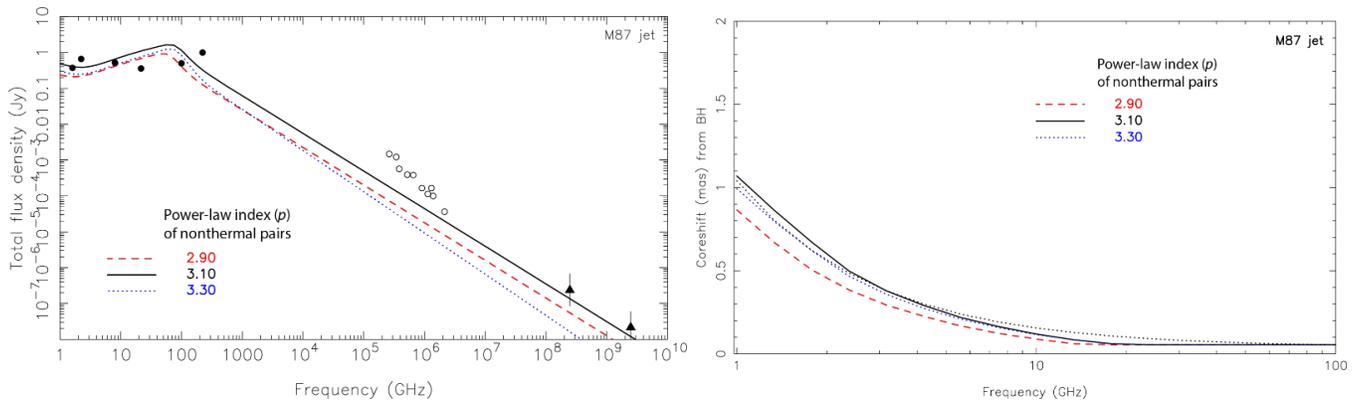

**Figure 16.** *Left:* Total SEDs for three different cases of the power-law index of nonthermal pairs, $p$. The black solid curve corresponds to the fiducial case (fig. 8), that is, $p = 3.1$, whereas the red dashed and blue dotted ones to $p = 2.9$ and $p = 3.3$, respectively. *Right:* Core shifts for the same three cases.

### 5.1. Proton contamination

We start with considering the value of $f_p$ below which a hadron contamination contributes significantly. For the present case of $p \approx 3$, the averaged Lorentz factor of nonthermal pairs becomes $\langle \gamma \rangle \approx 2\gamma_{\min}$, provided that $\gamma_{\max} \gg \gamma_{\min}$. Therefore, the proton's inertia affects the lepton density through equations (13) and (14) if $f_p$ is less than or nearly equals to $1 - (2\langle \gamma \rangle)/1836 \approx 0.998$ for $\gamma_{\min} \approx 1$. On the other hand, if $p \approx 2$, the harder lepton's energy distribution results in a greater lepton mass, $\langle \gamma \rangle \approx 10$, in the jet co-moving frame. Owing to such a heavy lepton mass, proton mass contributes to reduce the pair density significantly when $f_p < 0.99$ if $p \approx 2$.

In the present paper, we assume $p \approx 3$; thus, to show the difference clearly, we begin with examining the cases of $f_p = 0.99$ 0.90, and 0.50, which correspond to a contamination of protons at 1 %, 10 %, and 50 %, respectively, in number. In figure 17, we present the results for these three cases. The left panel shows that the flux density decreases enough when $f_p$ decreases to 0.99, as expected. For further smaller values of $f_p$, the flux density further decreases. This is because the lepton density, $n_*^{\mathrm{tot}}$, decreases through an increase of $U$ in equation (14) due to hadronic contribution in mass with decreasing $f_p$.

The right panel shows that the core shift decreases with decreasing $f_p$. This is because the decreased $n_*^{\mathrm{tot}}$ results in a reduced synchrotron absorption when the pair fraction decreases (or equivalently, when the proton fraction increases). Even when the proton contamination is only 1 % in number (i.e., when $f_p = 0.99$), both SED and core shift decrease at a non-negligible level from the case of a pure pair plasma, $f_p = 1.0$, because the fluid mass increases significantly by protons as stated above.

### 5.2. Normal-plasma-dominated jets

Next, we examine the case of hadron-dominated jets, adopting further smaller values of $f_p$. In the present section, we adopt $f_p = 0.90, 0.50$, and $0.10$, and compare the results with the case of a pure pair plasma, $f_p = 1.00$. Other parameters than $f_p$ are unchanged from the fiducial case (§ 3.3).

To examine how the spectral shape changes in radio frequencies, we adjust $B_{p,0}$ so that their flux densities may match in the optically thin frequency regime. It is achieved by adopting $B_{p,0} = 62.0$ G, $106.4$ G, and $183.0$ G, for $f_p = 0.90, 0.50$, and $0.10$, respectively, where $B_{p,0} = 25.0$ G for $f_p = 1.00$. In figure 18, the left panel shows their SEDs. The black solid curve (with $f_p = 1$) corresponds to the fiducial case discussed in § 4 for a pure pair plasma.

It follows from the left panel that the synchrotron spectrum becomes inverted in mm wavelengths if a normal-plasma dominated jet has sufficient magnetic-field strengths. The reasons are threefold. (1) A heavier mass of a normal-plasma dominated jet lowers the plasma density, when the Poynting flux is converted into the kinetic one. (2) The lower plasma density allows a greater magnetic-field strength so that the optically thin synchrotron flux may lie below the observed values between IR and X-ray frequencies. (3) The greater magnetic-field strength results in a higher peak frequency, as well as a stronger intensity of the synchrotron spectrum by the thermal emission at the jet base.

The right panel shows that the core shift and its gradient increase with increasing hadron fraction, $1 - f_p$, if we fix other parameters than $f_p$ and $B_{p,0}$. As a result, the core shift becomes too large when $f_p < 1$. Nevertheless, as figure 14 show, we can adjust the collimation power-law index $q$ so that the core shift may match observations while keeping the SED unchanged. Thus, we adjust $q$ and match both the SED and the core shift with observations, and present the result in figure 19. It follows that the flux densities increase between 30 and 1000 GHz (i.e., in mm-submm wavelengths) with decreasing the lepton fraction $f_p$, while the SED in other frequencies and the core shift unchanged. Thus, it is possible that we could infer the matter content, which is represented by $f_p$, by examining both the SED and the core shift. For this purpose, we should constrain $q$ independently.

To constrain $q$, we can use the jet shape. For example, at the projected distance 0.6 mas from the BH, the projected opening angle measured between the jet edges becomes 67.1°, 81.0°, 96.3°, and 118.2° for black solid, red dashed blue dotted, and green dash-dotted curves (or cases) in figure 19. On the other hand, the projected opening angle is reported to be between 55° and 60° at the same distance (Hada et al. 2016). Thus, within our present jet model, we can infer that the lepton-dominated jet (i.e., $f_p = 1.0$) is preferable, that the mass loading takes place at $r \approx 6.8M$ from the BH, that the magneticl field strength is about 25 G at this jet base, and that the power-law index of the magnetic-flux function $A_\varphi$ is $q \approx 0.75$.

We finally present the radio spectrum in figure 20 for the hadron-dominated case (green dash-dot-dot-dotted curve in fig. 19). The black solid curve shows the total spectrum, whereas the black dashed one does the thermal component. It follows that the thermal component appears prominently between 20 and 700 GHz when $\Theta = 2.0$ (or greater). This conclusion should be independently examined with ALMA between 230 and 690 GHz.

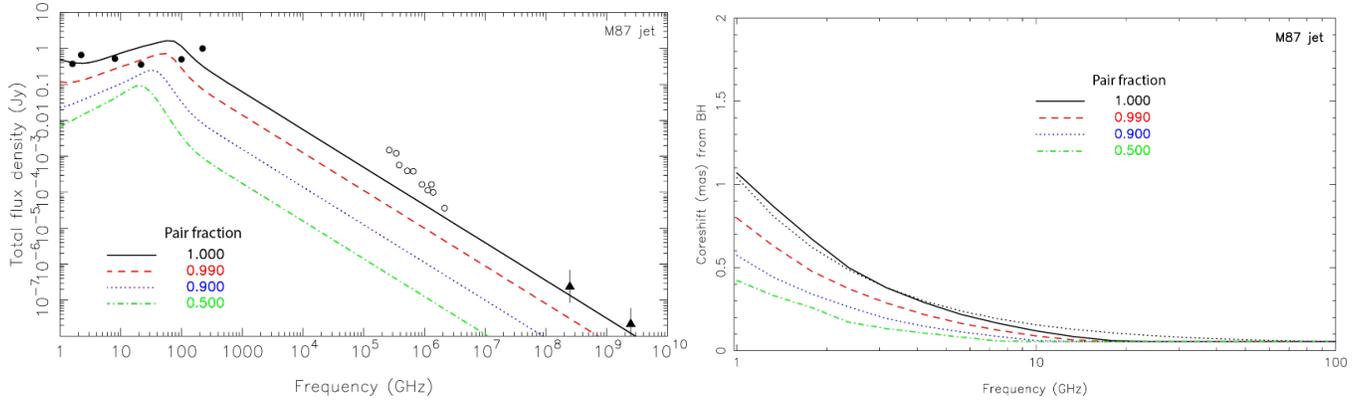

**Figure 17.** *Left:* Total SEDs for three different cases of pair fraction, $f_\mathrm{p}$. The black solid curve corresponds to the fiducial case (fig. 8), that is, $f_\mathrm{p} = 1.000$, whereas the red dashed, blue dotted, and green dash-dotted ones to $f_\mathrm{p} = 0.99$ $f_\mathrm{p} = 0.90$, and $f_\mathrm{p} = 0.50$, respectively. *Right:* Core shifts for the same three cases.

### 5.3. *Ring-like structure formed in hadronic jet*

Let us briefly investigate how the ring-like structure changes as a function of the lepton fraction, $f_\mathrm{p}$. In figure 21, we plot the expected VLBI map for three different values of $f_\mathrm{p}$. It follows that the ring-like structure is blurred with increasing proton fraction, $1-f_\mathrm{p}$ (or equivalently, with decreasing lepton fraction $f_\mathrm{p}$). Comparing with the VLBI map obtained at 86 GHz (fig. 1 of Lu et al. 2023), we find that the fiducial case ($f_\mathrm{p} = 1.0$, $B_\mathrm{p} = 25.0$ G, $q = 0.75$ with jet-launching altitide at $6.8R_\mathrm{S}$) reproduces the observations of SED, core shift, ring-like structure, the limb-brighted structure, and the jet opening angle at sub-parsec scales, at least qualitatively. It is noteworthy that $q = 0.75$ is consistent with the value inferred at larger distances (Asada & Nakamura 2012; Nakamura & Asada 2013), and gives a consistent width of the brightened limb at 0.6 mas and 0.8 mas from the core (Kim et al. 2018, H24).

### 6. DISCUSSION

In summary, we apply the `R-JET` code (H25), a post-processing radiative-transport code, to the M87 jet, utilizing the published GRMHD simulation results to constrain the angle-dependent energy extraction from a rapidly rotating BH (Tchekhovskoy et al. 2010) and the bulk Lorentz factor evolution along the jet (Mertens et al. 2016). Fitting the spectrum, the core shift, the radius of the ring-like structure, and the width of the brightened limbs with observations, we can constrain the jet parameters. The fitted results show that the power-law index of the magnetic flux function is $q = 0.75$, the jet-launching altitude is $6.8R_\mathrm{S}$, the magnetic-field strength is 25 G at this jet base, magnetization parameter evolves with altitude $r$ by $r^{-1}$, the jet is composed of a pure pair plasma, $f_\mathrm{p} = 1.0$, the power-law index of the nonthermal electrons is about $p = 3.1$, and the thermal lepton's temperature is about $kT = 2m_\mathrm{e}c^2$. For a hadronic jet, it is also possible to fit both the spectrum and the core shift; however, the predicted opening angle becomes much greater than the VLBI observations. Nevertheless, it is worth noting that the spectrum shows a hump between 30 and 1000 GHz due to the thermal emission from the inner-most part of the jet.

### 6.1. *Implication to ALMA observations*

Figure 20 shows the radio spectrum of a hadron-dominated jet when the lepton temperature is $\Theta_\mathrm{e} = 2$ at the jet base, $r = 6.8R_\mathrm{S}$. It follows that ALMA observations in this frequency range covers the frequency range in which the thermal component contributes significantly. For example, lower frequency range (ALMA bands 1) is important to find the deviation from a flat-top spectrum (below 20 GHz) to an inverted spectrum (above 20 GHz). Middle frequency range (bands 2–8) constrains the curvature of the thermal component. The higher frequency range (band 9–10) is important to constrain the frequency of the inflection point, which is formed between the thermal-component dominance (in lower frequencies) and the nonthermal-component dominance (in higher frequencies).

To closely look into the spectrum in sub-millimeter wavelengths, we highlight recent ALMA data in red in figure 20. The data points are 1.42 Jy at 86.3 GHz, 1.33 Jy at 221 GHz, and 0.97 Jy at 343 GHz (Peng et al. 2024; Goddi et al. 2021; Goddi & Carlos 2025). We set $\Theta_\mathrm{e} = 2.0$ at the jet base so that the flux densities in 22–43 GHz (black filled circles) and at 343 GHz (red filled circle) may be reproduced by the thermal component (i.e., the black dashed curve). The observed flux density at 343 GHz implies that the flat-top SED turns over around 200 GHz. During the high, flaring state, the open circles show that a flat-top SED is also obtained, although the flux density is not reported at 343 GHz. To explain this flat SED between 8 and 200 GHz, we considered various combinations of matter content (i.e., $f_{\rm p}$), lepton temperature ($\Theta_{\rm e}$), collimation index ($q$), and magnetic field strength ($B_{\rm p,0}$). However, we failed to reproduce the SED particularly between 86 and 211 GHz, keeping the core shift at observed values. Thus, we cannot conclude whether the sub-millimeter emission is thermal or non-thermal origin due to the discrepancy in 86–211 GHz.

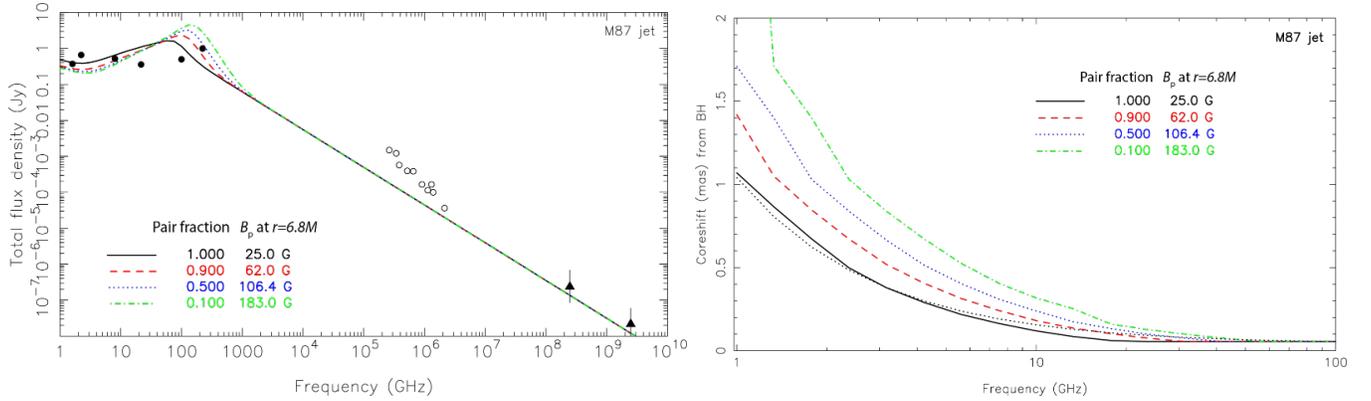

**Figure 18.** *Left:* Total SEDs for four different combinations of $f_{\rm p}$ and $B_{\rm p,0}$. The black solid curve corresponds to the fiducial case $(f_{\rm p},B_{\rm p,0})=(1.00,25.0{\rm G})$, whereas the red dashed, green dash-dotted, and blue dotted curves corresponds to $(0.90,62.0~{\rm G})$, $(0.50,106.4~{\rm G})$, and $(0.10, 183.0~{\rm G})$, respectively. *Right:* Core shifts for the same four cases.

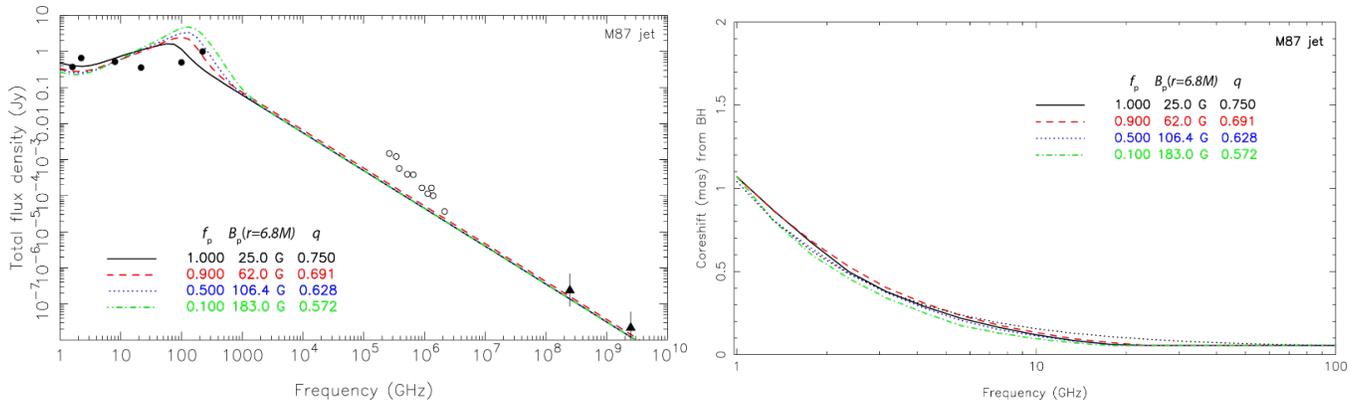

**Figure 19.** *Left:* Total SEDs for four different combinations of $f_{\rm p}$, $B_{\rm p,0}$, and $q$. The black solid curve corresponds to the fiducial case $(f_{\rm p},B_{\rm p,0},q)=(1.00,25.0{\rm G},0.750)$, whereas the red dashed, green dash-dotted, and blue dotted curves corresponds to $(0.90,62.0~{\rm G},0.691)$, $(0.50,106.4~{\rm G},0.628)$, and $(0.10,183.0~{\rm G},0.572)$, respectively. *Right:* Core shifts for the same four cases.

We consider that this discrepancy may arise because we assume a stationary and continuous jet. If the jet is nonstationary and discontinuous, the resultant high plasma density and magnetic field in the blob may lead to a higher turnover frequency by the opacity effect of synchrotron-self-absorption than a continuous jet. Accordingly, by superposing a few such (unresolved) blobs at different $r$'s, one could obtain a flat-top SED between e.g., 10 and 200 GHz. In another word, ALMA observations can strongly constrain the physical conditions in the jet launching region, such as the matter content, magnetic-field strength, collimation index, and the density contrast between a radio-emitting blob and other regions, because the photons in this frequency range are mostly emitted from the inner-most region of the jet. Since we assume a continuous jet in the present paper, we postpone the investigation of non-stationary, discontinuous jets to our subsequent paper.

### 6.2. *Constraining the position of shocks by the shape of VLBI jet*

In the present paper, we adjusted the magnetic-field collimation index, $q$, so that the predicted SED and core shift may match the observations. However, using the VLBI technique, Asada & Nakamura (2012) constrained the value of $q$ as 0.75 within the projected distance $z$ between $2\times10^2 R_{\rm S}$ and $10^5 R_{\rm S}$, which correspond to the de-projected distances $6.5\times10^2 R_{\rm S} < r < 3.2\times10^5$. What is more, Lu et al. (2023) revealed that the jet tends to be *more collimated* (than $q = 0.75$), quasi-cylindrical at de-projected distance $30R_S < r < 80R_S$ from the BH, where the core shift at infinite frequency ($\approx 19R_S$) is added to the axial distance presented in figure 3 of Lu et al. (2023) to derive the range of the de-projected distance *from the BH*.

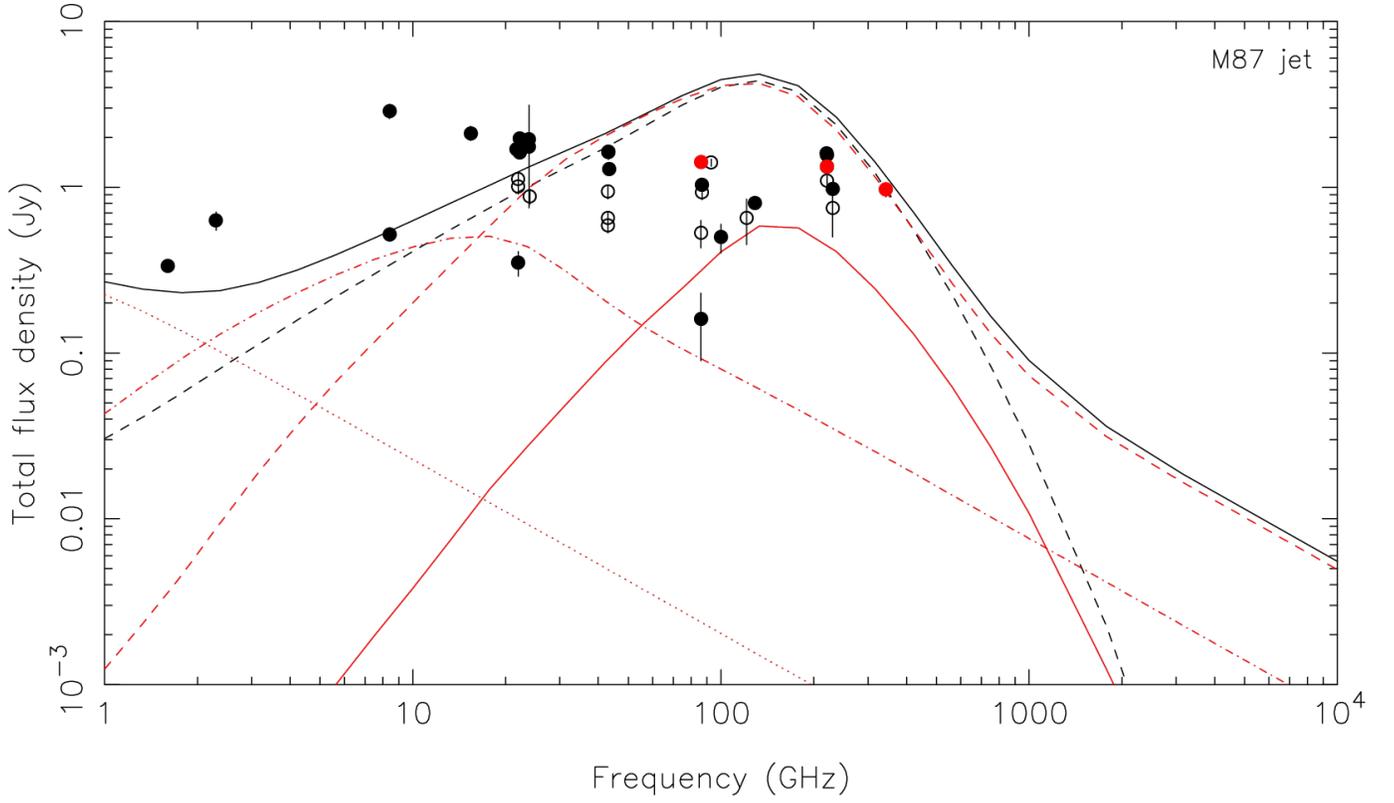

**Figure 20.** Expected SED of the M87 jet when $f_{\rm p} = 0.10$, $B_{\rm p,0} = 183.0$ G, and $q = 0.571$. Other parameters are the same as the fiducial case. The black solid curve represents the total SED, while the black dashed one does the thermal component. The red solid, red dahsed, red dash-dotted, and red dotted curves show the flux densities of the photons emitted within $6.8 < r/R_{\rm S} < 10^1$, $10^1 < r/R_{\rm S} < 10^2$, $10^2 < r/R_{\rm S} < 10^3$, and $10^3 < r/R_{\rm S} < 10^4$, respectively. Observational data points are common with figure 8.

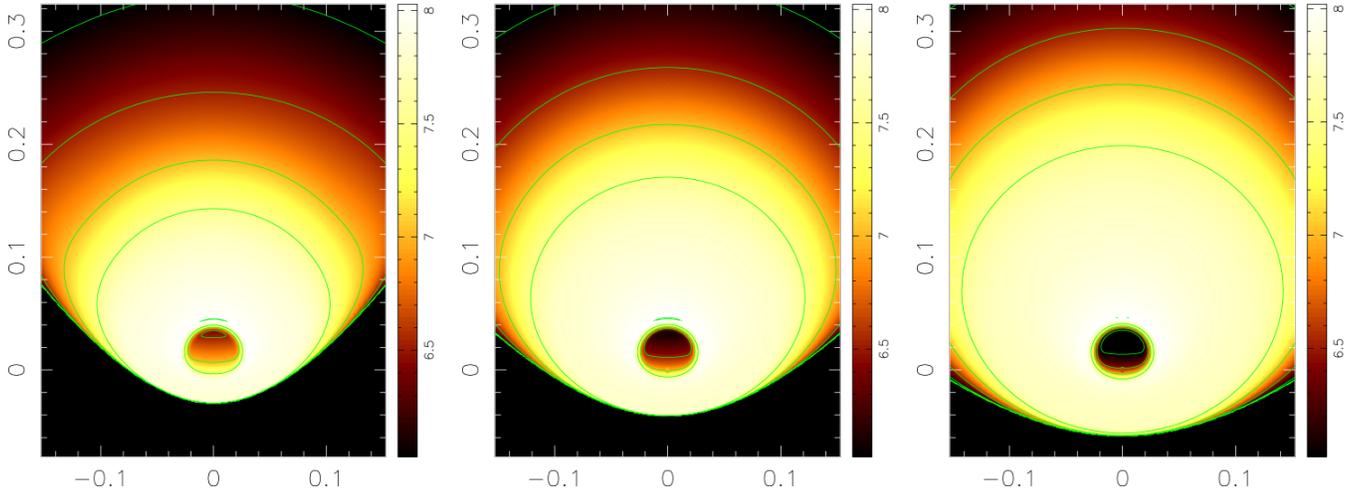

**Figure 21.** Surface brightness distribution at 86 GHz at the base of hadron-dominated jets. The left, middle, and right panels represent the results for $(f_{\rm p}, B_{\rm p}, q) =$ (0.9,62.0 G,0.691) (0.5,106.0 G,0.628), and (0.1,183.0 G,0.572), which correspond to the the red dashed, green dash-dotted, and blue dotted curves in figure 13, respectively. These three panels should be compared with the lepton-dominated, fiducial case (fig. 9; $f_{\rm p} = 1.0$, $B_{\rm p} = 25.0$ G, and $q = 0.750$). The peak brightness is 106.9 Jy mas$^{-2}$ 105.8 Jy mas$^{-2}$, 104.8 Jy mas$^{-2}$, from left to right.

The latter observation, which suggests a quasi-cylindrical jet, is particularly important, because it appears to contradict with our result, which points a *quasi-collimated* jet, $q = 0.75$. We consider that this discrepancy may be due to a time-dependent nature of the flow line geometry at the jet base. For example, during a very-low or quiescent phase, a small plasma density, and hence a weak electric current near the BH, will produce a magnetic field whose configuration is close to the Wald solution. Here, the Wald solution is a GR generalization of the cylindrical field geometry around a rotating BH, and is realized when the magnetic field is produced by an equatorial ring current flowing at large distances. On the other hand, during a flare phase, higher density plasmas near the BH will possess stronger toroidal electric currents, modifying the Wald solution to a parabolic or conical geometry in the poloidal plane (e.g., Tchekhovskoy et al. 2011). More research is, therefore, needed to examine the discrepancy of the flow-line geometry at the jet base. What is important is that the magnetic-field configuration in the inner-most region will be highly time-dependent, because the plasma density, and hence the electric currents will change significantly within short time scales there.

It is worth discussing if the present stationary model is consistent with such a time-dependent nature of the jet launching regions. In the present method, the assumption of a stationary jet is essential when we apply the energy conservation in equation (16). Thus, as long as the jet is magnetically dominated, equation (16) can be approximately satisfied, provided that the Poynting flux is nearly constant with time and that the energy loss (e.g., by radiation) is negligible. Accordingly, near the jet base where $\sigma \gg 1$, we adopt equation (16), although plasmas may contribute in energy in a time-dependent way at a negligible level. In the jet downstream, we consider that such inhomogeneity is smeared out and adopt equation (16).

On the other hand, as long as the flow-line shape is concerned, the collimation index $q$ can vary with $r$, because a stationary jet can change its flow-line shape at different positions. In fact, the R-JET code allows a spatially-changing $q = q(r)$, although we adopt a constant $q$ in this particular paper. Thus, as long as the varying time scale is long compared to the dynamical time scale, a spatially-changing $q(r)$ does not contradict with the assumption of a stationary and axisymmetric jet. For example, Hirotani et al. (2023) considered such a possibility and argued the constriction of the transverse width of the M87 jet on the projected, celestial plane.

### 6.3. *Application to other AGN jets*

In the case of the M87 jet, the thermal synchrotron emission from the jet base appears as a ring-like structure (figs. 9 & 21), because we are viewing the jet base almost face-on ($\theta_v \approx 18°$), and because the de-projected half opening angle of the jet ($> 30°$ within 0.1 mas) is greater than $\theta_v$ by virtue of the low-latitude concentration of the BZ flux (i.e., the limb brightening). If we apply the method described in the present paper to other AGN jets, it is, in principle, possible to constrain the jet-formation altitude and the composition. Namely, comparing the predicted SED, core shift, and ring radius with observations, we can constrain $f_p$ (matter content), $q$ (flow-line collimation), $B_p$ (magnetic-field strength), and the jet-forming altitude of an AGN jet.

In our subsequent papers, we will apply the R-JET code to other AGN jets, and further investigate the limb-brightened structure of nearby AGN jets, as well as their potential ring-like structure at the jet base.

When applying the R-JET code to various AGN jets systematically in the future, it is desirable to upgrade the present R-JET code. For example, as the first version, the present R-JET code can treat only the synchrotron process. However, to quantify the emission in higher frequencies (e.g., between IR and X-ray frequencies), it is essential to incorporate the synchrotron self-Compton process. What is more, incorporation of the polarized radiative transfer is also essential. Although the present R-JET code can treat only the total intensity, it is, indeed, designed so that an extension to a polarized case may be straightforward. Given the availability of EHT polarization data, we plan to incorporate polarization effects in the next version of the R-JET code.

The authors thank Hung-Yi Pu for valuable comments that improved the manuscript. The authors acknowledge support for the CompAS group under Theory from the Institute of Astronomy and Astrophysics, Academia Sinica (ASIAA), the Academia Sinica Grant AS-IAIA-114-M01, and the National Science and Technology Council (NSTC) in Taiwan through grants 113-2112-M-001-008-, the Greenland Telescope (GLT) grant 113-2124-M-001–008. The authors acknowledge access to high-performance facilities in ASIAA and thank the National Center for High-Performance Computing (NCHC) of the National Applied Research Laboratories (NARLabs) in Taiwan for providing computational and storage resources. This work utilized tools (BIWA GRPIC code) developed and maintained by the ASIAA CompAS group. This research has used the SAO/NASA Astrophysics Data System.

## REFERENCES

Abramowski, A., Acero, F., Aharonian, F., et al. 2012, ApJ, 746, 151, doi: 10.1088/0004-637X/746/2/151

Acciari, V. A., Aliu, E., Arlen, T., et al. 2009, Science, 325, 444, doi: 10.1126/science.1175406

Algaba, J. C., Baloković, M., Chandra, S., et al. 2024, A&A, 692, A140, doi: 10.1051/0004-6361/202450497

Aliu, E., Arlen, T., Aune, T., et al. 2012, ApJ, 746, 141, doi: 10.1088/0004-637X/746/2/141

Asada, K., & Nakamura, M. 2012, ApJL, 745, L28, doi: 10.1088/2041-8205/745/2/L28

Asada, K., Nakamura, M., Doi, A., Nagai, H., & Inoue, M. 2014, ApJL, 781, L2, doi: 10.1088/2041-8205/781/1/L2

Asada, K., Nakamura, M., & Pu, H.-Y. 2016, ApJ, 833, 56, doi: 10.3847/1538-4357/833/1/56

Beskin, V. S., Istomin, Y. N., & Parev, V. I. 1992, Soviet Ast., 36, 642

Beskin, V. S., Kniazev, F. A., & Chatterjee, K. 2023, MNRAS, 524, 4012, doi: 10.1093/mnras/stad2064

Beskin, V. S., & Nokhrina, E. E. 2009, MNRAS, 397, 1486, doi: 10.1111/j.1365-2966.2009.14964.x

Biretta, J. A., Sparks, W. B., & Macchetto, F. 1999, ApJ, 520, 621, doi: 10.1086/307499

Blakeslee, J. P., Jordán, A., Mei, S., et al. 2009, ApJ, 694, 556, doi: 10.1088/0004-637X/694/1/556

Blandford, R. D., & Levinson, A. 1995, ApJ, 441, 79, doi: 10.1086/175338

Blandford, R. D., & Payne, D. G. 1982, MNRAS, 199, 883, doi: 10.1093/mnras/199.4.883

Blandford, R. D., & Znajek, R. L. 1977, MNRAS, 179, 433, doi: 10.1093/mnras/179.3.433

Bransgrove, A., Ripperda, B., & Philippov, A. 2021, PhRvL, 127, 055101, doi: 10.1103/PhysRevLett.127.055101

Broderick, A. E., & Loeb, A. 2009, ApJ, 697, 1164, doi: 10.1088/0004-637X/697/2/1164

Broderick, A. E., & Tchekhovskoy, A. 2015, ApJ, 809, 97, doi: 10.1088/0004-637X/809/1/97

Celotti, A., & Fabian, A. C. 1993, MNRAS, 264, 228, doi: 10.1093/mnras/264.1.228

Chen, A. Y., & Yuan, Y. 2020, ApJ, 895, 121, doi: 10.3847/1538-4357/ab8c46

Chiueh, T., Li, Z.-Y., & Begelman, M. C. 1998, ApJ, 505, 835, doi: 10.1086/306209

Crinquand, B., Cerutti, B., Dubus, G., Parfrey, K., & Philippov, A. 2021, A&A, 650, A163, doi: 10.1051/0004-6361/202040158

Curtis, H. D. 1918, Publications of Lick Observatory, 13, 9

Doeleman, S. S., Fish, V. L., Schenck, D. E., et al. 2012, Science, 338, 355, doi: 10.1126/science.1224768

Dokuchaev, V. I. 2023, Astronomy, 2, 141, doi: 10.3390/astronomy2030010

EHT MWL Science Working Group, Algaba, J. C., Anczarski, J., et al. 2021, ApJL, 911, L11, doi: 10.3847/2041-8213/abef71

Event Horizon Telescope Collaboration, Akiyama, K., Alberdi, A., et al. 2019a, ApJL, 875, L6, doi: 10.3847/2041-8213/ab1141

—. 2019b, ApJL, 875, L1, doi: 10.3847/2041-8213/ab0ec7

—. 2019c, ApJL, 875, L6, doi: 10.3847/2041-8213/ab1141

Feng, J., & Wu, Q. 2017, MNRAS, 470, 612, doi: 10.1093/mnras/stx1283

Frolova, V. A., Nokhrina, E. E., & Pashchenko, I. N. 2023, MNRAS, 523, 887, doi: 10.1093/mnras/stad1381

Fuentes, A., Gómez, J. L., Martí, J. M., & Perucho, M. 2018, ApJ, 860, 121, doi: 10.3847/1538-4357/aac091

Gebhardt, K., Adams, J., Richstone, D., et al. 2011, ApJ, 729, 119, doi: 10.1088/0004-637X/729/2/119

Giovannini, G., Feretti, L., & Comoretto, G. 1990, ApJ, 358, 159, doi: 10.1086/168970

Goddi, C., & Carlos, D. F. 2025, arXiv e-prints, arXiv:2505.10181, doi: 10.48550/arXiv.2505.10181

Goddi, C., Martí-Vidal, I., Messias, H., et al. 2021, ApJL, 910, L14, doi: 10.3847/2041-8213/abee6a

Hada, K., Doi, A., Kino, M., et al. 2011, Nature, 477, 185, doi: 10.1038/nature10387

Hada, K., Kino, M., Nagai, H., et al. 2012, ApJ, 760, 52, doi: 10.1088/0004-637X/760/1/52

Hada, K., Kino, M., Doi, A., et al. 2013, ApJ, 775, 70, doi: 10.1088/0004-637X/775/1/70

—. 2016, ApJ, 817, 131, doi: 10.3847/0004-637X/817/2/131

Harris, D. E., Cheung, C. C., Stawarz, Ł., Biretta, J. A., & Perlman, E. S. 2009, ApJ, 699, 305, doi: 10.1088/0004-637X/699/1/305

Hirotani, K., & Okamoto, I. 1998, ApJ, 497, 563, doi: 10.1086/305479

Hirotani, K., & Pu, H.-Y. 2016, ApJ, 818, 50, doi: 10.3847/0004-637X/818/1/50

Hirotani, K., Shang, H., Krasnopolsky, R., & Nishikawa, K. 2023, ApJ, 943, 164, doi: 10.3847/1538-4357/aca8b0

—. 2024, ApJ, 965, 50, doi: 10.3847/1538-4357/ad28c7

—. 2025, ApJ, 984, 16, doi: 10.3847/1538-4357/adbaf2

Junor, W., & Biretta, J. A. 1995, AJ, 109, 500, doi: 10.1086/117295

Junor, W., Biretta, J. A., & Livio, M. 1999, Nature, 401, 891, doi: 10.1038/44780

Kim, J. Y., Krichbaum, T. P., Lu, R. S., et al. 2018, A&A, 616, A188, doi: 10.1051/0004-6361/201832921

Kino, M., Takahara, F., Hada, K., et al. 2015, ApJ, 803, 30, doi: 10.1088/0004-637X/803/1/30

Kino, M., Takahara, F., Hada, K., & Doi, A. 2014, ApJ, 786, 5, doi: 10.1088/0004-637X/786/1/5

Kisaka, S., Levinson, A., & Toma, K. 2020, ApJ, 902, 80, doi: 10.3847/1538-4357/abb46c

Koide, S., Shibata, K., Kudoh, T., & Meier, D. L. 2002, Science, 295, 1688, doi: 10.1126/science.1068240

Komissarov, S. S., Vlahakis, N., Königl, A., & Barkov, M. V. 2009, MNRAS, 394, 1182, doi: 10.1111/j.1365-2966.2009.14410.x

Konigl, A. 1981, ApJ, 243, 700, doi: 10.1086/158638

Kovalev, Y. Y., Lister, M. L., Homan, D. C., & Kellermann, K. I. 2007, ApJL, 668, L27, doi: 10.1086/522603

Kramer, J. A., & MacDonald, N. R. 2021, A&A, 656, A143, doi: 10.1051/0004-6361/202141454

Kutkin, A. M., Sokolovsky, K. V., Lisakov, M. M., et al. 2014, MNRAS, 437, 3396, doi: 10.1093/mnras/stt2133

Lee, S.-S., Lobanov, A. P., Krichbaum, T. P., et al. 2008, AJ, 136, 159, doi: 10.1088/0004-6256/136/1/159

Levinson, A., & Rieger, F. 2011, ApJ, 730, 123, doi: 10.1088/0004-637X/730/2/123

Li, Y.-R., Yuan, Y.-F., Wang, J.-M., Wang, J.-C., & Zhang, S. 2009, ApJ, 699, 513, doi: 10.1088/0004-637X/699/1/513

Liepold, E. R., Ma, C.-P., & Walsh, J. L. 2023, ApJL, 945, L35, doi: 10.3847/2041-8213/acbbcf

Lobanov, A. P. 1998, A&A, 330, 79. https://arxiv.org/abs/astro-ph/9712132

Lonsdale, C. J., Doeleman, S. S., & Phillips, R. B. 1998, AJ, 116, 8, doi: 10.1086/300417

Lu, R.-S., Asada, K., Krichbaum, T. P., et al. 2023, Nature, 616, 686, doi: 10.1038/s41586-023-05843-w

Ly, C., Walker, R. C., & Junor, W. 2007, ApJ, 660, 200, doi: 10.1086/512846

Lyubarsky, Y. 2009, ApJ, 698, 1570, doi: 10.1088/0004-637X/698/2/1570

McKinney, J. C. 2005, ApJL, 630, L5, doi: 10.1086/468184

Mertens, F., Lobanov, A. P., Walker, R. C., & Hardee, P. E. 2016, A&A, 595, A54, doi: 10.1051/0004-6361/201628829

Morabito, D. D., Niell, A. E., Preston, R. A., et al. 1986, AJ, 91, 1038, doi: 10.1086/114080

Morabito, D. D., Preston, R. A., & Jauncey, D. L. 1988, AJ, 95, 1037, doi: 10.1086/114700

Nakamura, M., & Asada, K. 2013, ApJ, 775, 118, doi: 10.1088/0004-637X/775/2/118

Nakamura, M., Asada, K., Hada, K., et al. 2018, ApJ, 868, 146, doi: 10.3847/1538-4357/aaeb2d

Neronov, A., & Aharonian, F. A. 2007, ApJ, 671, 85, doi: 10.1086/522199

Nokhrina, E. E., & Pushkarev, A. B. 2024, MNRAS, 528, 2523, doi: 10.1093/mnras/stae179

Ogihara, T., Takahashi, K., & Toma, K. 2019, ApJ, 877, 19, doi: 10.3847/1538-4357/ab1909

Parfrey, K., Philippov, A., & Cerutti, B. 2019, PhRvL, 122, 035101, doi: 10.1103/PhysRevLett.122.035101

Peng, S., Lu, R.-S., Goddi, C., et al. 2024, ApJ, 975, 103, doi: 10.3847/1538-4357/ad7c41

Prieto, M. A., Fernández-Ontiveros, J. A., Markoff, S., Espada, D., & González-Martín, O. 2016, MNRAS, 457, 3801, doi: 10.1093/mnras/stw166

Reynolds, C. S., Fabian, A. C., Celotti, A., & Rees, M. J. 1996, MNRAS, 283, 873, doi: 10.1093/mnras/283.3.873

Ricci, L., Boccardi, B., Nokhrina, E., et al. 2022, A&A, 664, A166, doi: 10.1051/0004-6361/202243958

Ro, H., Kino, M., Sohn, B. W., et al. 2023, A&A, 673, A159, doi: 10.1051/0004-6361/202142988

Rybicki, G. B., & Lightman, A. P. 1986, Radiative Processes in Astrophysics

Sokolovsky, K. V., Kovalev, Y. Y., Pushkarev, A. B., & Lobanov, A. P. 2011, A&A, 532, A38, doi: 10.1051/0004-6361/201016072

Takahashi, K., Toma, K., Kino, M., Nakamura, M., & Hada, K. 2018, ApJ, 868, 82, doi: 10.3847/1538-4357/aae832

Tchekhovskoy, A., Narayan, R., & McKinney, J. C. 2010, ApJ, 711, 50, doi: 10.1088/0004-637X/711/1/50

—. 2011, MNRAS, 418, L79, doi: 10.1111/j.1745-3933.2011.01147.x

Vlahakis, N. 2004, ApJ, 600, 324, doi: 10.1086/379701

Walsh, J. L., Barth, A. J., Ho, L. C., & Sarzi, M. 2013, ApJ, 770, 86, doi: 10.1088/0004-637X/770/2/86

Weber, E. J., & Davis, Leverett, J. 1967, ApJ, 148, 217, doi: 10.1086/149138